\documentclass[sn-mathphys-num, referee]{sn-jnl}
\usepackage{amsmath,amsfonts}
\usepackage{graphicx}    
\usepackage{bm}          
\usepackage[utf8]{inputenc}
\usepackage[T1]{fontenc}
\usepackage{mathptmx}
\usepackage{gensymb}
\usepackage{physics}
\usepackage{xcolor}
\usepackage{xfrac}
\usepackage{float}
\usepackage{mathtools}
\usepackage{lipsum}

\begin{document}

\title[Vector beams in ENZ metamaterials]{Longitudinal field controls vector vortex beams in anisotropic epsilon-near-zero metamaterials
} 

\author*[1]{\fnm{Vittorio} \sur{Aita}}\email{vittorio.aita@kcl.ac.uk}
\equalcont{These authors contributed equally to this work.}

\author*[1]{\fnm{Diane J.} \sur{Roth}}\email{diane.roth@kcl.ac.uk}
\equalcont{These authors contributed equally to this work.}

\author[1]{\fnm{Anastasiia} \sur{Zaleska}}\email{anastasiia.zaleska@kcl.ac.uk}
\author[1]{\fnm{Alexey V.} \sur{Krasavin}}\email{alexey.krasavin@kcl.ac.uk}
\author[1]{\fnm{Luke H.} \sur{Nicholls}}\email{luke.nicholls@kcl.ac.uk}
\author[1,2]{\fnm{Mykyta} \sur{Shevchenko}} \email{mykyta.shevchenko@npl.co.uk}
\author[1]{\fnm{Francisco J.} \sur{Rodríguez-Fortuño}}\email{francisco.rodriguez\_fortuno@kcl.ac.uk}
\author[1]{\fnm{Anatoly V.} \sur{Zayats}}\email{a.zayats@kcl.ac.uk}

\affil[1]{\orgdiv{Department of Physics and London Centre for Nanotechnology}, \orgname{King's College London}, \orgaddress{\street{Strand}, \city{London}, \postcode{WC2R 2LS}, \country{UK}}}
\affil[2]{\orgname{National Physical Laboratory}, \orgaddress{\street{Hampton Road}, \city{Teddington}, \postcode{TW11 0LW}, \country{UK}}}

\date{\today}

\abstract{Structured light plays an important role in metrology, optical trapping and manipulation, communications, quantum technologies, nonlinear optics and provides a rich playground for addressing new optical phenomena. Here we demonstrate a novel approach for manipulating vector vortex beams carrying longitudinal field components using metamaterials with extreme anisotropy. Implementing vectorial spectroscopy, we show that the propagation of complex beams with inhomogeneous polarisation is strongly affected by the interplay of the metamaterial anisotropy with the transverse and longitudinal field structure of the beam. This phenomenon is especially pronounced in the epsilon-near-zero regime, exclusively realised for light polarised along the metamaterial optical axis, strongly influencing the interaction of longitudinal fields with the metamaterial. The requirements on the balance between the transverse and longitudinal fields to maintain a polarisation singularity at the beam axis allow control of the beam modal content, filtering diffraction effects and tailoring spatial polarisation distribution. The proposed approach offers important capabilities for wavefront shaping as well as local spatial polarisation engineering. The understanding of the interaction of vector beams with metamaterials opens new opportunities for applications in microscopy, information encoding, biochemical sensing and quantum technologies.}

\maketitle 

\section{Introduction}\label{sec:Introduction}
The ability to precisely tailor and manipulate the properties of optical wavefronts, such as amplitude, phase and polarisation state, is paramount for the development of various applications including imaging, metrology, optical communications as well as biochemical and quantum technologies~\cite{bliokh2023roadmap, Willner2021, Forbes2021, Li2024, shen2019optical}. In this respect, complex beam profiling, which exploits the vectorial nature of electromagnetic waves becomes increasingly important~\cite{Zhan09,shen2019optical,shen2024optical,shen2023topological}. Vector vortex beams have emerged as a new tool for wavefront shaping, thanks to their unique properties offering new degrees of freedom in engineering spatially nonuniform polarisation distributions~\cite{sugic2021particle, bliokh2023roadmap,singh2023robust}. These beams have characteristic doughnut-shaped intensity profiles, featuring polarisation singularities with a strong longitudinal (directed along the wavevector) field. 
Two of the most common types of vector beams with cylindrical symmetry are the so-called radial and azimuthal beams with radial and azimuthal orientation of the electric field across the beam profile, respectively. Due to the continuity of the electric field required by Maxwell's equations, radial beams exhibit a strong longitudinal electric field component, centred on the beam optical axis. Conversely, a strong longitudinal magnetic field component together with a purely transverse electric field is found in azimuthal beams. A precise balance between the transverse and longitudinal components of the field is needed in such inhomogeneously polarised beams in order to allow their propagation~\cite{aita2022enhancement, Afanasev2023}.

Typically, longitudinal components are present in guided or strongly focused electromagnetic fields and responsible for their transverse spin~\cite{Aiello2015,Aiello_2016,Vernon2024}. This unusual property, which may be present even in unpolarised light~\cite{eismann2021transverse}, is an origin of many counterintuitive effects, such as directional scattering, unidirectional coupling to waveguides, lateral optical forces, and photonic topological structures~\cite{Luxmoore2013,Junge2013,Fortuño2013,Petersen2014,Bliokh2015,Rodríguez-Fortuño2015, Kapitanova2014, leFeber2015,Neugebauer2015,Picardi2019}. The longitudinal field of vector beams has a profound impact on their scattering properties when they interact with matter. In the case of individual nanoparticles, it induces dipoles in the direction of the beam propagation and, therefore, is scattered in the directions normal to the incident wave vector~\cite{krasavin2018generalization}. 

Different techniques have been developed for the generation and manipulation of vector vortex beams~\cite{naidoo2016controlled,rosales2018review,karimi2010polarization}, including the use of static phase plates and liquid crystal cells~\cite{liu2017generation,brasselet2009optical,rumala2013tunable,chen2015generation}, as well as spatial light modulators ~\cite{liu2018highly,perez2017demand}. It is also possible to generate vector beams from a non-monochromatic source, exploiting the total internal reflection from a glass cone~\cite{radwell2016achromatic}. Metamaterials and metasurfaces with spatially varying distribution of meta-atoms, which imprint the required phase distribution on an incoming optical field, have also been recently shown to provide an excellent platform for the generation and manipulation of structured beams~\cite{yue2016vector,chen2016geometric,ding2020focused}, potentially easing their integration into photonic devices. 

In this paper, we experimentally and theoretically demonstrate that strongly anisotropic metamaterials--here realised as a two-dimensional array of gold nanorods--can be used to modify the polarisation structure and modal content of vector beams, exploiting the drastic difference in the metamaterial response to transverse and longitudinal components of the electric field. When vector vortex beams propagate in such media, the balance between longitudinal and transverse electric field components is broken by anisotropic absorption, requiring the redistribution of the energy between these field components, in order to sustain the propagation. This is especially pronounced in the epsilon-near-zero regime of the metamaterial dispersion for focused radial beams at normal incidence propagating along the optical axis since the effect on the longitudinal field component is the strongest. Under these conditions, the evolution of the energy distribution between the transverse and longitudinal field components of the vector beam results in the filtering of high-order modes of the beam induced by the focusing and, therefore, diffraction suppression. The understanding of the interaction of vector beams with anisotropic metamaterials opens up new opportunities for applications of polarised beams and nanostructured materials in optical trapping, information encoding and biochemical sensing.

\section{Results}\label{sec:results}

\subsection{Longitudinal field spectroscopy}
\begin{figure}[!ht]
    \centering
    \includegraphics[width =  \linewidth]{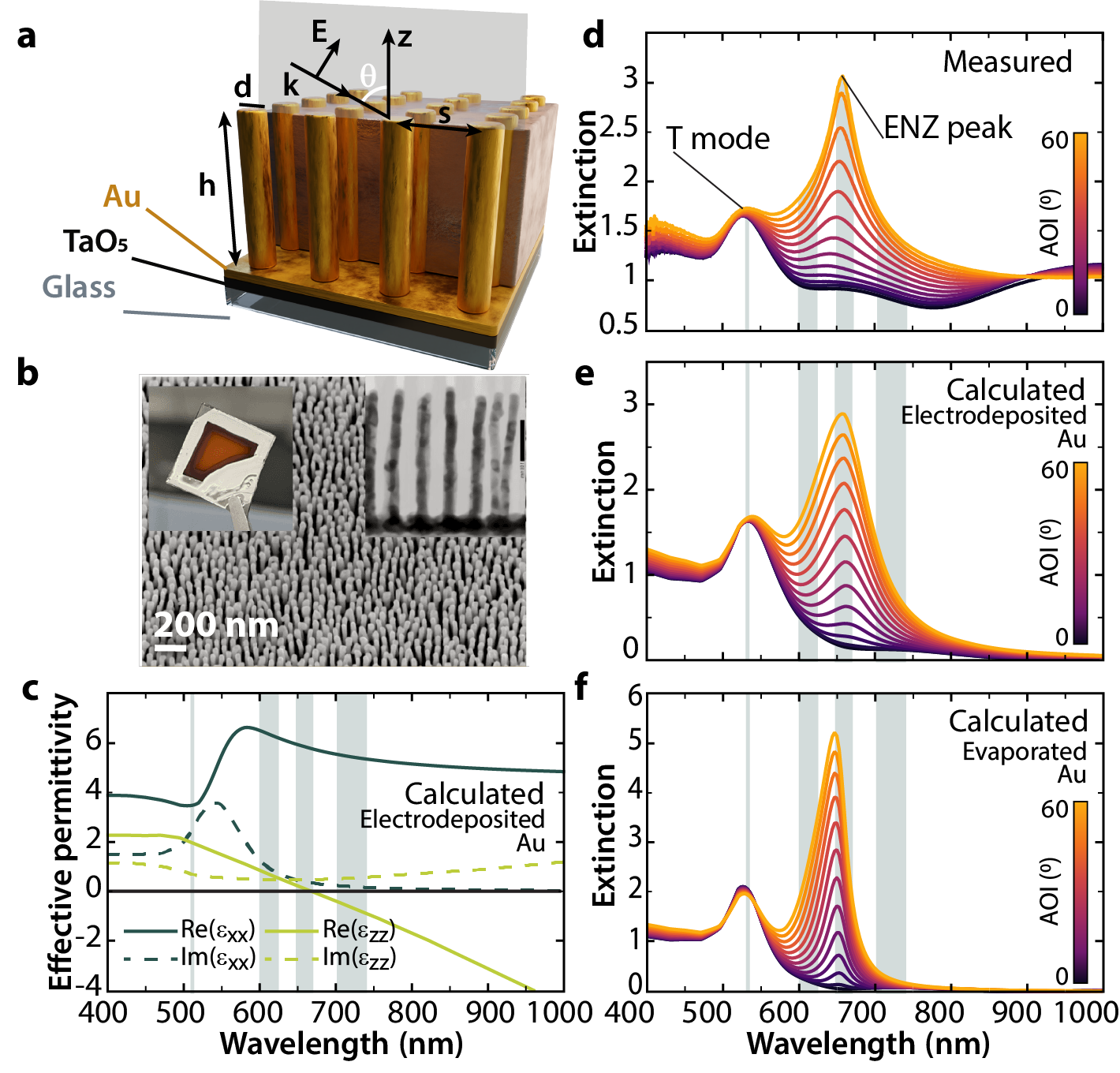}
    \caption{\textbf{Optical properties of the uniaxial metamaterial.} (a) Schematics and (b) SEM image of the metamaterial, with the insets showing a photograph of the sample and the TEM cross-section of the nanorods. (c) Real and imaginary parts of the effective permittivity tensor for the considered metamaterial (d=34~nm, h=400~nm, s=69~nm). (d--f) Extinction spectra obtained under $p$-polarised plane wave illumination for different angles of incidence, as indicated by the colourbar~\cite{crameri2020misuse}: (d) experiment, (e,f) effective medium model for the permittivity of (e) electrodeposited and (d) evaporated gold. Shaded areas in (c--f) represent the bandwidth of the filters used for wavelength selection in the experiments in Sections 2.2 and 2.3.}
    \label{fig:Figure1}
\end{figure}

Vector vortex beams can be described by the paraxial Helmholtz equation in cylindrical coordinates, seeking a vectorial solution of radial or azimuthal symmetry, respectively~\cite{gori1987bessel}. Formally, the solutions obtained are Bessel-Gauss beams, but they are often approximated with the Laguerre-Gauss (LG) modes LG$_{\rm{\ell p}}$, where $\ell$ and p represent the azimuthal and radial mode numbers. Within the Jones formalism, the transverse field (in the  $x-y$ plane normal to the wavevector) of radial and azimuthal vectorial modes can be presented as a superposition of two orthogonally polarised circular scalar vortices~\cite{maurer2007tailoring,moreno2010decomposition}:
\begin{subequations}
\begin{align}
    \va{E}_{\rm Rad} &= \rm{LG}_{10}\vu{n}_R + \rm{LG}_{-10}\vu{n}_L\\
    \va{E}_{\rm Azi} &= \rm{LG}_{10}\vu{n}_R + e^{i\pi}\rm{LG}_{-10}\vu{n}_L\,,
\end{align}
\end{subequations}\label{eq:RadAzi_LG}
where the unit vectors $\vu{n}_{R,L}$ indicate the right- and left-handed circular polarisation states. This representation makes use of two scalar optical vortices $\rm{LG}_{\pm10}$ possessing phase singularities of opposite topological charges ($\ell = \pm1$). The phase of their superposition naturally has a null topological charge, and the centre of the intensity distribution represents a singularity in the polarisation distribution. The local polarisation orientation of radial and azimuthal beams is such that the transverse electric field vector (Eq.~\ref{eq:RadAzi_LG}) describes a whole $2\pi$ rotation along a closed loop around the beam centre, which can define a winding number, similarly to the topological charge for scalar vortices~\cite{kotlyar2022index}. Such spatial variation of the transverse polarisation enforces a constraint on the longitudinal field that the beam may carry. In the absence of sources, the three-dimensional electric field must be divergence-free, so that the structure of the transverse field requires the presence of a longitudinal field component to satisfy Maxwell's equations~\cite{aita2022enhancement}: $\mathbf{\nabla}\cdot\mathbf{E}=0$. In particular, the transverse component of an azimuthal beam is sufficient to make it divergence-free, so that this polarisation state can not sustain any longitudinal field. On the contrary, a radial beam needs a nonzero longitudinal component ${E}^{z}_{\rm Rad}\neq$0 to lower its total divergence to zero, even under the paraxial approximation~\cite{forbes2021relevance}: $\mathbf{\nabla}\cdot\mathbf{E} = \mathbf{\nabla}_{\rm T}\cdot\mathbf{E}^{\rm Rad}_{\rm T} + ik E^{\rm Rad}_z = 0$, where $k$ is the wavenumber. This divergence-free condition needs to be satisfied as the beam propagates and interacts with a medium. 

To exploit this unique property for controlling vector beams, we employed a uniaxial anisotropic metamaterial based on plasmonic nanorod assembly (Fig.~\ref{fig:Figure1}a,b)~\cite{Roth2024}. It consists of an array of gold nanorods embedded in an alumina matrix supported by a glass substrate (see Methods for the details of the fabrication). In the absence of pronounced nonlocal effects, the metamaterial optical properties can be described within a local effective medium theory using the Maxwell Garnett approximation~(see Methods)~\cite{elser2006nanowire}. The metamaterial behaves as a highly anisotropic uniaxial medium with the optical axis parallel to the nanorods and permittivity tensor given by $\pmb \epsilon^{\rm eff} = \mathrm{diag}\left[\epsilon_{xx},\epsilon_{yy}=\epsilon_{xx},\epsilon_{zz}\right]$. For the chosen structural parameters, $\epsilon_{xx,yy}$ are always positive~(Fig.~\ref{fig:Figure1}c), while $\epsilon_{zz}$ changes sign at a wavelength around 660~nm, marking the so-called epsilon-near-zero (ENZ) region that plays a key role in a broad range of applications~\cite{Roth2024}. 

The extinction spectra of the metamaterial, measured under $p$-polarised plane wave illumination, show a typical behaviour of this type of nanostructure with two extinction peaks, related to the metamaterial strong anisotropy (Fig.~\ref{fig:Figure1}d). The one at a wavelength around $\lambda_{\rm T}\approx530$~nm is associated with the dipolar resonance ($\rm T$-mode) excited with a field perpendicular to the nanorod axis. 
The extinction peak located at a wavelength around $\lambda_{\rm ENZ}\approx 660$~nm is related to the ENZ behaviour of the metamaterial for light polarised along the nanorods and, therefore, observed only at oblique incidence under plane wave illumination. The effective model (see Methods) well describes the experimental spectra considering the optical properties of the electrodeposited gold corrected for the electron mean-free-path of 8~nm (compared to 10.8~nm of bulk gold)~\cite{johnson1972optical,etchegoin2006analytic,etchegoinerratum}. The gold permittivity controls the quality factor of the ENZ resonance in the extinction (Fig.~\ref{fig:Figure1}e). 

Given these properties of the metamaterial, the ENZ-related extinction cannot be harnessed by plane wave illumination at normal incidence, as it does not provide a field component along the optical axis. However, a linearly or circularly polarised Gaussian beam can acquire a longitudinal field upon focusing. At the same time, longitudinal field is intrinsic for radial and other vector vortex beams even within the paraxial approximation~\cite{forbes2021relevance} and can be further enhanced by focusing. 

\begin{figure}[!ht]
    \centering
    \includegraphics[width =  \linewidth]{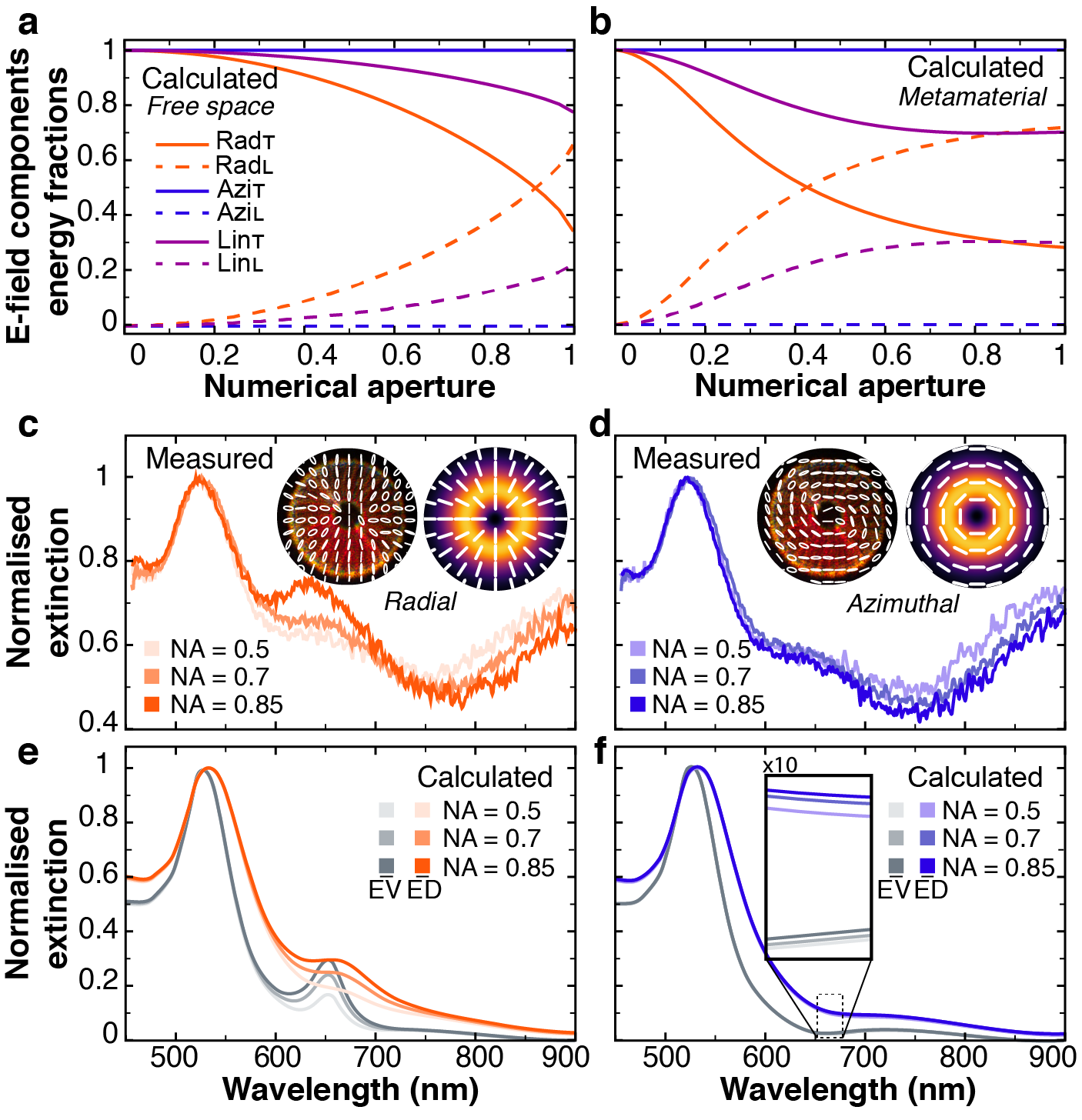}
    \caption{\textbf{Vector beam spectroscopy.} (a,b) Relative energy stored in the (solid lines) transverse and (dashed) longitudinal components of the electric field with respect to the total energy contained in the beam for propagation in (a) free space and (b) the metamaterial (the metamaterial parameters are as in Fig.~\ref{fig:Figure1}c) for (orange) radial, (purple) linear, and (blue) azimuthal polarisations, calculated in the focal plane ($z=0$) with the metamaterial slab being centred on it. (c--f) Extinction spectra obtained with focused vector beams at normal incidence for NA of 0.5, 0.7, and 0.85 (increasing with the colour tone): (c,d) measured and (e,f) calculated extinction spectra of the metamaterial for (c,e) radial and (d,f) azimuthal beams and for (coloured) electrodeposited and (greyscale) evaporated gold. Extinction spectra are individually normalised to their own maxima for direct comparison. 
    }
    \label{fig:Figure2}
\end{figure}

The choice of the initial state of polarisation of the beam before focusing allows for tailoring of the energy exchange between the transverse and longitudinal fields, governed by the requirements of divergence-free electric fields from Maxwell's equations (Fig.~\ref{fig:Figure2}a). 
The relative contribution of the longitudinal component to the total intensity generally increases with the numerical aperture (NA) of the focusing. For a radial beam focused in free space, under the low focusing conditions (NA = 0.1) the longitudinal intensity corresponds to less than $1\%$ of the overall intensity in the focal plane ($z=0$), whereas the contribution of the longitudinal field overcomes the one of the transverse field intensity at NA$\approx$0.9. As can be expected, an azimuthal beam has a null longitudinal field regardless of the NA, while a linearly polarised beam possesses a nonzero $E_z$ although never as strong as that of the radial beam.

When focused inside a metamaterial, a monochromatic radial beam at a wavelength corresponding to the ENZ condition, acquires the stronger longitudinal field exceeding the transverse field intensity at the much lower NA = 0.4 (Fig.~\ref{fig:Figure2}b) due to the metamaterial anisotropy, suggesting the possibility to generate a relatively stronger longitudinal component with less focused beams. Similar observations can be made for the linearly polarised beam, while for the azimuthally polarised beam, the longitudinal component keeps the zero value. This effect is strongly wavelength-dependent because of the drastically different interaction between the longitudinal field and the metamaterial in the different dispersion regimes. While the anisotropy increases in the hyperbolic regime compared to the elliptic one, only the ENZ regime produces a strong damping of the longitudinal field~\cite{aita2024PRB}.

This behaviour can be directly visualised in the extinction spectra of the radial and azimuthal beams (Fig.~\ref{fig:Figure1}). Vector beam spectroscopy (see Methods) reveals significant extinction of the radial beam at normal incidence, which increases with the stronger focusing (Fig.~\ref{fig:Figure2}c,d). Therefore, in the ENZ regime, the metamaterial sensitivity to the angle of incidence shown with plane waves can be transformed to a polarisation structure sensitivity in the case of strongly focused vector vortex beams. In the experimental extinction spectra obtained with radial beam illumination, the ratio of the magnitudes of the ENZ-peak to the T-peak is more than double its theoretical estimate (Fig.~\ref{fig:Figure2}e,f).  Comparing the extinction spectra for the electrodeposited and evaporated gold, one can see that while the widths of the extinction peaks are considerably smaller in the latter case, their magnitudes remain essentially the same, which was not observed for the plane wave case (Fig.~\ref{fig:Figure2}e,f). These observations may be an indication of the interconnected nature of the transverse and longitudinal fields in the vector beam, which is absent in the case of the focused Gaussian beam illumination.

The differences between the propagation in free space and through the metamaterial can be understood as a shift of the balance from transverse to the longitudinal components of the field in the metamaterial case (Supplementary Fig.~\ref{fig:FigS1}) due to a significant increase of the latter in the ENZ regime, in accordance with the corresponding boundary conditions at the air-metamaterial interface. The strongly modified field distributions are the result of the transformation of the radial eigenmode over the full vector beam eigenmode set caused by the presence of the metamaterial layer and the focusing. In the reference case of glass, a strong longitudinal field component is clearly visible in the centre of the beam, as well as the presence of higher order modes, as shown by the external rings appearing in the pattern, as a consequence of the focusing/diffraction only. On the other hand, the propagation of the beam through the metamaterial differently influences the transverse and longitudinal field components of the beam, depending on the wavelength. The absorption of the longitudinal field component increases as the wavelength gets closer to $\lambda_{\rm ENZ}$, and consequently, the central peak almost completely disappears compared to the non-absorbing case (glass) at the same wavelength. Additionally, the intensity distribution of the transverse field component is structurally modified. Comparing the propagation in glass and the metamaterial at $\lambda_{\rm ENZ}$, one can observe the external rings disappear and the doughnut-shaped distribution gets smaller in the metamaterial. This indicates that the interaction of focused vector beams with strongly anisotropic medium described above modifies the beam modal content and its polarisation state through the tailoring of the longitudinal field as discussed in the following sections. 

\subsection{Modal content filtering}

\begin{figure}
    \centering
    \includegraphics[width =  \linewidth]{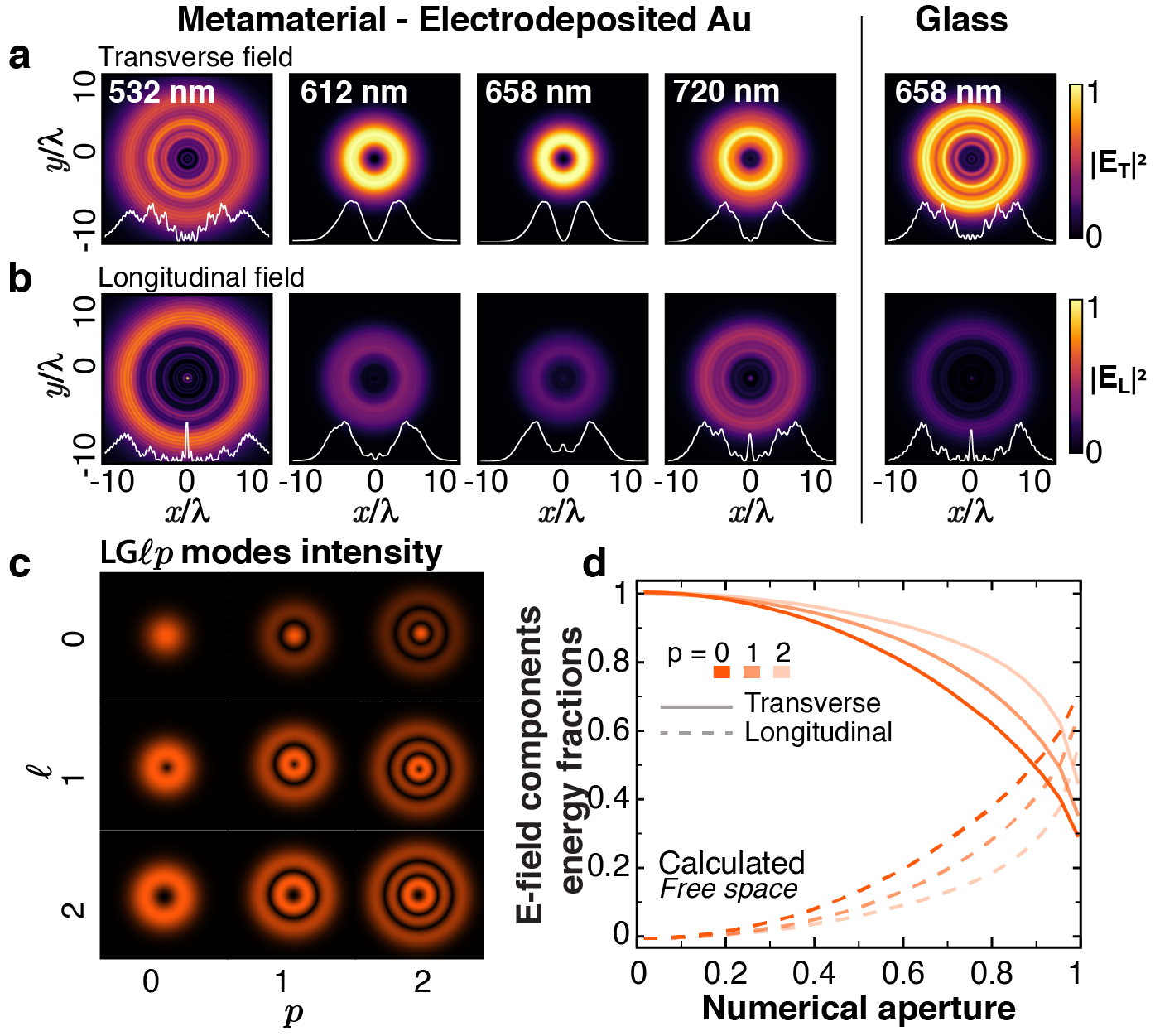}
    \caption{\textbf{Modal content of a tightly focused radial beam.} Theoretical intensity distribution of (a) transverse and (b) longitudinal field of a tightly focused (NA = 0.85) radial beam propagating through (four left columns) the metamaterial and (right column) glass for the same wavelengths as in the experiment. Cross-sections of the intensity profiles along the beam axis are shown as the inserts. (c) Intensity distributions of LG modes with $\ell$=0--2 and p=0--2. (d) Relative energy stored in (solid lines) transverse and (dashed lines) longitudinal components of the electric field of radial beams obtained as a superposition of ${\rm LG}_{\pm1\rm p}$ modes, with p = 0,1,2 (orange, decreasing colour tone) with respect to their total energy for propagation in free space (cf. Fig.~\ref{fig:Figure2}a,b).}
    \label{fig:Figure3}
\end{figure}
Taking into account the observations on the divergence-free nature of the electric field in the absence of sources and the role of focusing in the generation of a strong longitudinal field (Fig.~\ref{fig:Figure2}), a tightly focused radial beam ensures strong interaction of the longitudinal field with the metamaterial in the ENZ spectral range. At the same time, strong focusing leads also to appearance of strong diffraction effects in the propagating beam, resulting in higher-order LG modes contributing to the beam profile. A comparison of the focal plane intensity distributions of a focused radial beam propagating through the glass and the metamaterial~\cite{aita2024PRB} (see Methods for the details of the simulations) shows the strong influence of the metamaterial on the beam profile, which also depends on the wavelength (Fig.~\ref{fig:Figure3}). 

The LG$_{\pm1,0}$ modes describing paraxial radially polarised beam (Eq.~\ref{eq:RadAzi_LG}(a)) have the characteristic doughnut shape of the intensity profiles. Changing the modes quantum numbers affects the mode intensity distribution: higher $\lvert\ell\rvert$ creates wider doughnuts, $p\neq0$ causes the appearance of outer rings (Fig.~\ref{fig:Figure3}c). Importantly, modes ${\rm LG}_{\pm1p}$ with $p>0$ carry a strong longitudinal field upon focusing, similar to the radial beam obtained with $p=0$. Due to the diffraction effects, the intensity distributions of the focal plane fields can be represented as a superposition of LG modes of higher orders with a nonzero longitudinal field component. 

In the case of propagation in glass and in the metamaterial for wavelengths away from the ENZ region, a strong longitudinal peak is present in the centre of the focused radial beam~(Fig.~\ref{fig:Figure3}). Close to the ENZ wavelength, two effects can be noticed: both transverse and longitudinal intensity profiles become smoother and the strength of the central peak of the longitudinal field is strongly damped. As a result of the interplay between the modes contained in the field structure and the metamaterial anisotropy, the contribution of higher-order modes to the electric field is reduced. The absorption of the longitudinal field by the metamaterial requires redistribution of the energy from the transverse field to the longitudinal one in order to maintain the beam propagation, given the relation between these fields prescribed by Maxwell's equations. As a result, the metamaterial effectively filters the higher-order modes out of the beam, leaving the mode LG$_{10}$ dominating over the others. 

To confirm the above phenomenological description, the propagation of a focused radial beam and the intensity distributions of its polarisation components have been simulated at the wavelength $\lambda_{\rm ENZ}$ for free space, glass and the metamaterial sample in order to retrieve the modal content for different focusing conditions (Supplementary Fig.~\ref{fig:FigS2}). 
These distributions were then used in a fitting procedure in order to determine the beam modal content (see Methods). The experimental results were measured and analysed using the same procedure.

\begin{figure}
    \centering
    \includegraphics[width =  \linewidth]{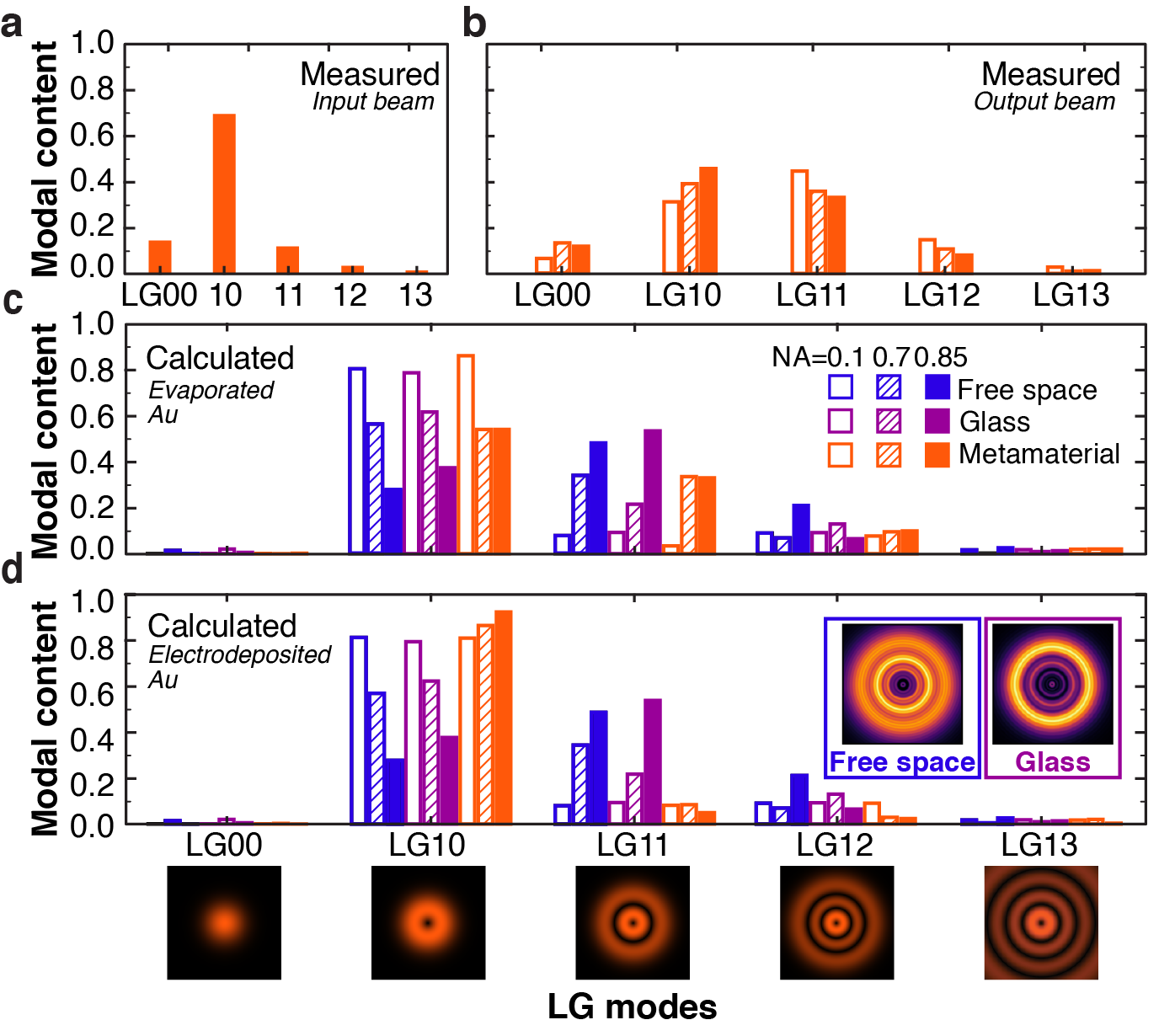}
    \caption{\textbf{Diffraction suppression of a tightly focused radial beam.} (a,b) Measured modal content of (a) input unfocused and (b) output focused radial beam propagating through the metamaterial. (c,d) Simulated modal content of a focused radial beam in (blue) free space, (purple) glass and (orange) the metamaterial for (c) evaporated and (d) electrodeposited gold permittivity: (empty bars) NA = 0.1, (patterned) NA = 0.7 and (filled) NA = 0.85. The theoretical spatial intensity distributions of the main LG modes are also shown. Insets in (d) show the intensity distributions obtained for a radial beam propagating in free space (blue frame) and glass (purple frame) under tight focusing (NA = 0.85), calculated in the focal plane. Each mode contribution is normalised to the sum of all the mode amplitudes contained in each set.}
    \label{fig:Figure4}
\end{figure}

Both experimental and modelling data confirm that the modal content of a paraxial radial beam predominantly comprises the mode LG$_{10}$ (cf. Fig.~\ref{fig:Figure4}a and Fig.~\ref{fig:Figure4}c,d). The modifications of the mode content due to focusing are clearly seen for propagation in free space (Fig.\ref{fig:Figure4}c,d), for which the increase in NA shows suppression of a LG$_{10}$ mode in favour of higher-order modes (mainly LG$_{11}$ and LG$_{12}$). Similar behaviour is observed also for propagation through a glass slab with small differences which can be ascribed to the reflection effects (Fig.~\ref{fig:Figure4}c,d). At the same time, the strong interaction with the metamaterial in the ENZ regime is clearly seen in the modal content of the beam (Fig.~\ref{fig:Figure4}b--d). The relative contribution of higher modes decreases while the LG$_{10}$ mode amplitude increases. This effect is more pronounced for stronger focusing. Experimental results are in good agreement with the theoretical predictions (Fig.~\ref{fig:Figure4}b). 

Hence, high anisotropy and ENZ behaviour of the metamaterial can be exploited to efficiently filter higher-order modes generated because of diffraction and achieve nearly diffractionless focusing of vector beams carrying longitudinal field. Please note that the quality of a vector beam is crucial for the experimental observation and applications of both extinction and modal filtering because imperfections of the polarisation state (\emph{e.g.}, local ellipticity) can drastically lower the strength of the longitudinal field and the requirements on its presence in the beam profile~\cite{aita2022enhancement}.

\subsection{Polarisation filtering}\label{sec:lowNA}

\begin{figure}
    \centering
    \includegraphics[width =  \linewidth]{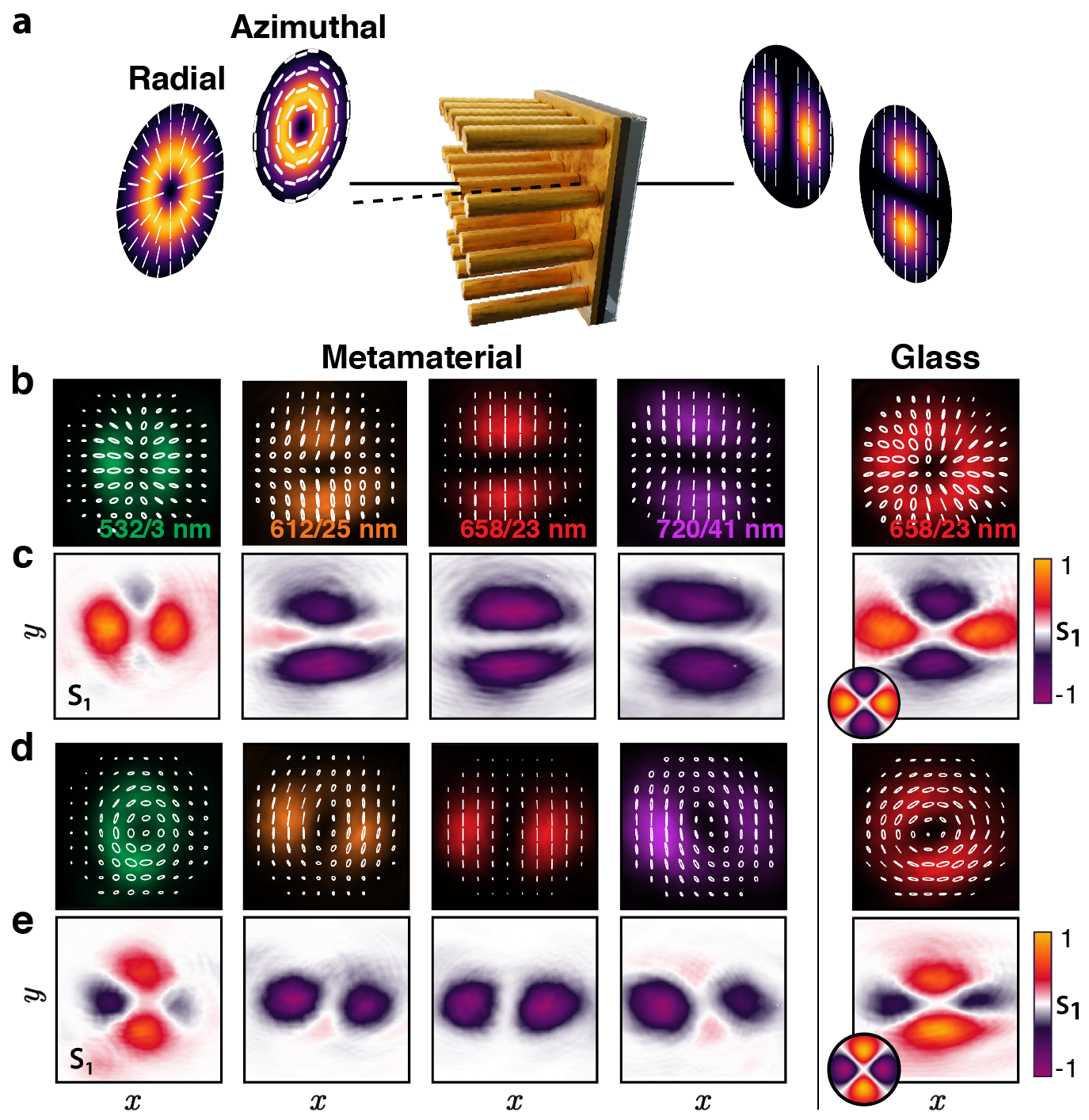}
    \caption{\textbf{Metamaterial as a linear polariser.} (a) Schematics of the polarisation measurements. (b,d) Retrieved polarisation of the transmitted light superimposed onto the beam intensity profile and (c,e) spatial distributions of the Stokes parameter $S_1$ for (b,c) radial and (d,e) azimuthal beams. Insets in the last column show the calculated distributions of $S_1$ for propagation in glass. Colours of the intensity profiles indicate the illumination wavelength: (green) $532\pm 1.5$~nm, (orange) $612 \pm 12.5$~nm, (red) $658 \pm 11.5$~nm and (purple) $ 720 \pm 20.5$~nm. The numerical aperture is NA = 0.1 and the angle of incidence is $60\degree$.}
    \label{fig:Figure5}
\end{figure}

In the case of weak focusing, the longitudinal field may not be strong enough to influence the modal structure significantly, but the beam intensity and vector structure may be modified by exploiting the metamaterial anisotropy. The interaction and modification of a paraxial vector beam can be enhanced at oblique incidence when the transverse field component also interacts with the ENZ resonance. In this case, the uniaxial metamaterial can act as a linear polariser or a waveplate, depending on wavelength and angle of incidence~\cite{Ginzburg2013}. In an isotropic medium (glass), the radial beam does not experience appreciable modification to its intensity profile and polarisation, as should be expected (Fig.~\ref{fig:Figure5}). However, for propagation through the metamaterial, the polarisation state and intensity profile of the beam are strongly modified and exhibit a strong wavelength dependence. The polarisation of the beam is globally converted to linear, while its intensity profile is changed from a doughnut shape to a two-lobed shape with increasing efficiency while approaching the ENZ region. The effect is similar to the transmission of a vector beam through a linear polariser with the intensity strongly reduced in those points where the original local polarisation is parallel to the nanorod axis. 
The spatial distribution of polarisation shows a clear transition from radial and azimuthal polarisations to a linear one. With the reduction of the angle of incidence, the observed effect is gradually reduced until it disappears at normal incidence. 

The behaviour of the metamaterial as a chromatic linear polariser can be quantitatively characterised by its extinction ratio and bandwidth. The polarisation extinction as the ratio of maximum to minimum transmission for two orthogonal linear polarisations can be obtained from the spectra in Fig.~\ref{fig:Figure1}. For an angle of incidence of 60$\degree$, a polarisation extinction ratio of 144:1 is observed with a bandwidth of approximately 20~nm, centred around $\lambda_{\rm ENZ}$~(Supplementary Figure~\ref{fig:FigS3}). The theoretical value corresponds to almost 150000:1 with about 10~nm bandwidth. This illustrates the potential of gold nanorods as a platform for ultrathin ($\leq$400~nm) wavelength selective polarisation devices. 

\section{Discussion}
We experimentally and theoretically investigated the interaction of vector beams with an extremely anisotropic uniaxial metamaterial also exhibiting ENZ properties for specific directions of polarisation. We showed that with different choices of polarisation states and focusing conditions, the longitudinal field of vector beams can be tailored to achieve strong coupling with the anisotropic ENZ behaviour of the metamaterial, even at normal incidence along the optical axis of the metamaterial. The anisotropic absorption of the transverse and longitudinal fields and the requirements on the electric field divergence lead to suppression of the diffraction of a radial beam via manipulation of its longitudinal field. 
In the paraxial (weak focusing) regime in which the longitudinal field is weak, the anisotropy of the metamaterial can be taken advantage of at oblique incidence, exploiting anisotropic absorption of the transverse field, which results in the modifications of the intensity and vector structure of the beams, since in these conditions the metamaterial acts as a linear polariser with a narrow bandwidth.  
The spectral position of the anisotropic ENZ regime, important for the interaction with a longitudinal field, can be tailored throughout visible and infrared regions at the fabrication stage. Thus, the interaction of vector vortex beams with anisotropic metamaterials demonstrates the great versatility of such a platform for designed vectorial wavefront shaping, which is of significant interest for various types of applications including optical communications, microscopy and metrology. 


\section{Methods}
\subsection{Sample fabrication}
The nanorod-based metamaterial with controlled geometrical parameters was fabricated by electrochemical deposition of gold into a porous alumina matrix on a supporting substrate~\cite{evans2006growth,Zaleska2024}. The substrate presents a multilayered structure composed of a glass slide (0.7~mm in thickness), a 10-nm-thick tantalum pentoxide (${\rm Ta}_{2}{\rm O}_{5}$) adhesive layer and an 8-nm-thick gold film acting as a weakly conducting electrode for the subsequent Au nanorod electrodeposition. An aluminium film (370~nm in thickness) is deposited onto the substrate by magnetron sputtering and subjected to a two-step anodisation process in  0.3~M sulphuric acid (${\rm H}_{2}{\rm SO}_{4}$) at 25~V to produce a porous anodic aluminium oxide (AAO) template. The geometrical parameters of the nanopores, such as diameter, separation and ordering, are controlled by the anodisation conditions (\emph{i.e.}, the choice of the electrolyte acid, its temperature and anodisation voltage) and the duration of pre-deposition chemical etching using a 30~mM sodium hydroxide solution (NaOH). Gold electrodeposition is performed with a three-electrode system using a non-cyanide gold plating solution. The length of the gold nanorods is controlled by the electrodeposition time.

\subsection{Effective Medium Theory modelling}

Optical properties of the nanorod-based metamaterial are derived using a local effective medium theory based on the Maxwell Garnett approximation~\cite{elser2006nanowire} which describes the behaviour of a uniaxial crystal. The effective permittivity tensor can then be written in terms of its principal components as follows $\pmb \epsilon^{\rm eff} = \mathrm{diag}\left[\epsilon_{xx},\epsilon_{yy}=\epsilon_{xx},\epsilon_{zz}\right]$, where the in-plane ($xy$-directions) and out-of-plane ($z$-direction) components of the effective dielectric permittivity are expressed as
\begin{subequations}\label{eq:eps_all}
\begin{align} \label{eq:eps_bot}
\epsilon_{xx} = \epsilon_{yy} &= \epsilon_{\rm h} \, \frac{\left(1+b\right)\epsilon_{\rm m}+\left(1-b\right)\epsilon_{\rm h}}{\left(1-b\right)\epsilon_{\rm m}+\left(1+b\right)\epsilon_{\rm h}} \,, \\
\label{eps_z}
\epsilon_{zz} &= b\,\epsilon_{\rm m}+\left(1-b\right)\epsilon_{\rm h}\,,
\end{align}
\end{subequations}
where $b = \pi\left(r / s\right)^2$ is the metal filling factor with $r$ being
the radius of the nanorods and $s$ being the periodicity of the array, in the assumption of a square lattice. The permittivities of metal and the host medium are denoted as $\epsilon_{\rm m}$ and $\epsilon_{\rm h}$, respectively. This model is valid within the effective medium approximation, i.e. away from the first Brillouin zone edge of the nanorod array. The effective permeability of a gold nanorod metamaterial is unity. The permittivity of the host medium (aluminium oxide, ${\rm Al}_{2}{\rm O}_{3}$) is taken from Ref.~\cite{dodge1986alumina} and $\epsilon_{\rm m}$ of evaporated gold from Ref.~\cite{johnson1972optical}. For electrodeposited gold, the permittivity was corrected by setting the restricted electron mean free path to 8~nm~\cite{lissberger1974optical}, which takes into account the variations due to the electrodeposition fabrication process.

The plot of the anisotropic effective permittivity components (Fig.~\ref{fig:Figure1}d) shows that the metamaterial supports two distinct dispersion regimes: elliptic with $\epsilon_{xx,yy}>0, \, \epsilon_{zz}>0$ and hyperbolic with $\epsilon_{xx,yy}>0, \, \epsilon_{zz}<0$. These two regimes are separated by the so-called epsilon-near-zero region at around a wavelength of $\lambda_{ENZ}=660$~nm, where Re$\epsilon_{zz}\approx$0. From the effective permittivity, the extinction spectra can be calculated using the transfer matrix model~\cite{Born_Wolf1999}. 

%

\subsection{Generation of vector vortex beams}

\begin{figure}
    \centering
    \includegraphics[width =  \linewidth]{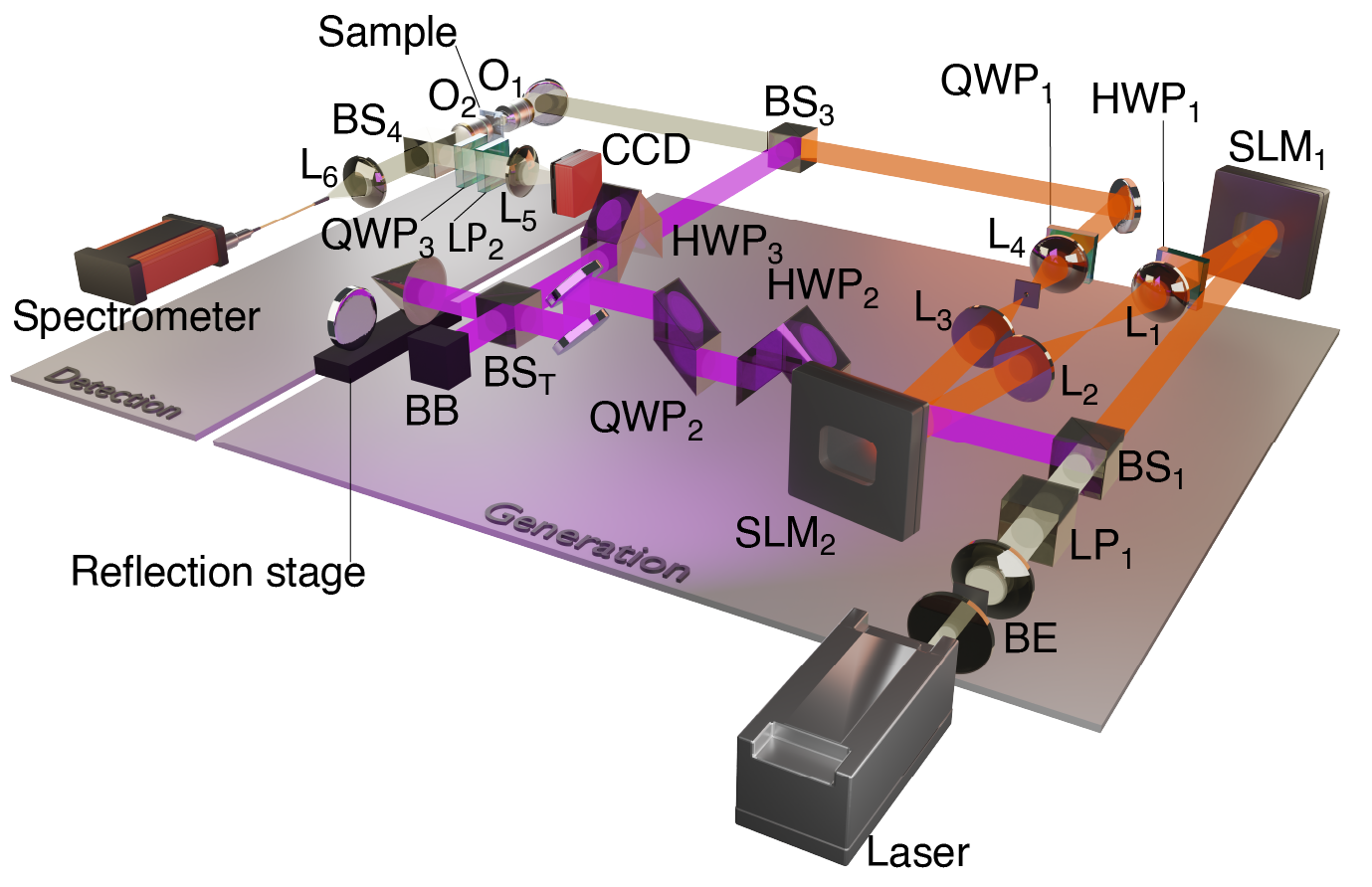}
    \caption{\textbf{Experimental setup.} Schematic representation of the experimental setup for beam preparation (generation stage) and optical characterisation (detection stage). The generation stage comprises two branches corresponding to (orange) monochromatic and (purple) broadband vector beams generation. 
    Optical components are labelled as follows: BE - beam expander, LP - linear polariser, BS - beam splitter, SLM - spatial light modulator, L - lens, HWP - half waveplate, QWP - quarter waveplate, O - objective, BB - beam blocker. The focal lengths of the lenses are: $f_1 = f_2 = 150$~mm, $f_3 = f_4 = 200$~mm, $f_5 = f_6 = 200$~mm. The sample is mounted on a customised stage allowing full translation and rotation.}
    \label{fig:Figure6}
\end{figure}

The output beam of a fibre-coupled supercontinuum laser (Fianium, SC-450-2) is expanded to the desired size by a pair of converging lenses in a 2-$f$ configuration (BE) and spatially filtered with a 100~$\mu$m pinhole to get a clean beam profile (Fig.~\ref{fig:Figure6}). The initial polarisation of the beam is fixed to be horizontal (parallel to the table, $x$ direction) by a Glan-Taylor prism (LP1), after which it is then sent to the generation stage of the setup. The setup consists of two branches: the (orange) monochromatic and (purple) broadband generation paths, both coupled to the same detection stage through BS3. 
The designed beam is brought to the back aperture of an objective (O1) with NA selected among 0.1, 0.5, 0.7 and 0.85, and focused on the sample. Light is collected by a second objective of NA = 0.9 (O2), which is kept the same for each illumination objective. The transmitted beam is sent to a 50-50 beamsplitter cube (BS4) either towards a CCD for image detection and polarisation analysis in the case of single wavelength measurements, or coupled to a spectrometer via a multimodal fibre in the case of broadband measurements. 

{\bf Broadband generation.} 
The horizontally polarised white beam first propagates through a series of 
double and single Fresnel Rhombs, used respectively as half- and quarter-waveplates (HWP2, QWP2) for full control of the scalar polarisation of the beam. The beam undergoes a double internal reflection inside a conical mirror of right-angle aperture which causes a spatially varying polarisation distribution~\cite{radwell2016achromatic}. The obtained vector vortex beams can be controlled to a certain extent by modifying the input polarisation (rotation of HWP2) and rotating the additional double Fresnel Rhomb (HWP3) in the output. With a careful orientation of these components, the polarisation state can be changed between radial and azimuthal. The orientation of HWP2 and HWP3 can be further modified to obtain linearly or circularly polarised light. In this instance, the reflection stage, which consists of a translation stage mounting the cone mirror and a flat circular mirror, is translated in order to introduce a flat mirror in place of a conical one. 

{\bf Monochromatic generation.} For monochromatic generation, the chosen wavelength is filtered from the supercontinuum spectrum before the beam is expanded. This is done with coloured filters of selected bandwidths and central wavelengths. 
The monochromatic branch is based on two spatial light modulators (PLUTO2 - NIR011, HOLOEYE), which allow versatility in the type of beams that can be generated~\cite{Fu2016,Wang2019,rosales2017shape}. Vector vortex beams are obtained as the superposition of scalar vortices of topological charge $\pm1$.
Both SLMs are oriented such that their modulation axis is aligned with the $x$-axis, which corresponds to the polarisation direction of the source. The first SLM provides a vortex phase term of $\exp\left(i\ell_1\phi\right)$, resulting in the Jones vector $\ket{H,\ell_1}$ for the polarisation state. This SLM also reflects the beam, giving it an additional $\pi$ phase shift, modifying the beam polarisation to $\ket{H,-\ell_{1}}$. A 4-$f$ telescope is mounted between the two SLMs to reduce the phase variations that would be induced by the propagation between the modulators. Here, a broadband half waveplate (HWP1), whose axis is oriented at an angle of $\theta_H = \pm\pi/8$ with the horizontal axis, rotates the incoming field by an angle of $\pi/4$, changing its polarisation to a diagonal state ($\ket{D}$ or $\ket{A}$). At this stage, the modulation given by SLM1 is stored in both the horizontal and vertical components of the field. When the beam reaches SLM2, its vertical field component is not modulated while the horizontal one acquires a new phase vortex factor $\exp(i\ell_2\phi)$. Taking also into account the reflection phase shift due to SLM2, the Jones vector after SLM2 becomes
\begin{equation*}
    \ket{{\rm out}_{\,\rm SLM2}} = \frac{1}{\sqrt{2}}\left(\ket{H, \ell_1-\ell_2}\pm\ket{V,\ell_1}\right)\ ,
\end{equation*}
where the choice in sign depends on whether HWP1 axis angle is set to $+$ or $-$ $\pi/8$. A broadband quarter waveplate (QWP1) with the axis oriented at $\theta_Q = \pm\pi/4$ with respect to the $x$-direction is then used to transform the $\ket{H}$ and $\ket{V}$ polarisation states into $\ket{R}$ and $\ket{L}$, eventually giving as output Jones vector
\begin{equation*}
    \ket{F} = \frac{1}{\sqrt{2}}\left(\ket{R,\ell_1-\ell_2} \pm e^{i \Phi}\ket{L,\ell_1}\right)\ .
\end{equation*}
Here, $\Phi$ is a constant phase term representing a phase difference between the two circular vortices that can be tuned by carefully choosing the phase masks on both SLMs. Lastly, a pinhole is placed between L3 and L4 so that the beam is spatially filtered after the modulation procedure has been performed. Using this method, vector vortex beams with an arbitrary topological charge, as well as standard linearly or circularly polarised beams can be generated by a simple change of the phase masks reproduced on the modulators and the waveplates orientation. The choice of the phase masks on SLM1 and SLM2 is dependent on the chosen operational wavelength.

\subsection{Vector field spectroscopy}
For the extinction measurements under vector vortex beam illumination, the incident beam was defined using the broadband generation technique described above. The beam was directed to the detection branch after transmission through the metamaterial and collected by the spectrometer.
To obtain extinction spectra, the average of transmitted signal over a set of 50 repeated measurements, each with the same exposure time of 8~ms was used. The transmitted signal through a glass slab of the same thickness as the substrate of the metamaterial was used as the reference signal.

Extinction measurements for a varied angle of incidence of $p$-polarised plane waves were taken on a similar setup, with the source changed to a high-power halogen lamp and the sample mounted on a rotation motor to allow precise control over the angle of incidence.

\subsection{Polarimetry}\label{sec:polarimetry}
Polarimetry characterisation is performed using monochromatic vector beams, prepared as described above. To obtain the Stokes parameters of the transmitted beam, four images of the beam intensity are taken for LP2 rotated by the angle of $0, \pi/4, \pi/2$ and $3\pi/4$ with respect to the $x$-axis, while two additional images are taken with QWP3 fixed at $\pi/4$ and LP2 either at $0$ or $\pi/2$ to create a circular analyser. With the measured transmitted intensity through the metamaterial denoted with $I(\alpha,\theta)$, where $\alpha$ is the angle of the fast axis of the quarter waveplate (marked as $\oslash$ in the case when the quarter waveplate is not present) and $\theta$ is the angle of the linear polariser, the Stokes parameters are calculated as
\begin{subequations}
\begin{align}
    S_{0} &= I\left(\oslash,0\degree\right) + I\left(\oslash,90\degree\right) \\
    S_{1} &= I\left(\oslash,0\degree\right) - I\left(\oslash,90\degree\right) \\
    S_{2} &= I\left(\oslash,45\degree\right) - I\left(\oslash,135\degree\right) \\
    S_{3} &= I\left(0\degree,45\degree\right) - I\left(0\degree,135\degree\right).
\end{align}
\end{subequations}
To recover the spatial distribution of the polarisation states of the transmitted light, additional images of the beam are collected after propagation through LP2. The analyser is rotated by steps of $20\degree$ in the range of $\theta = 0$ to $2\pi$. 
For an electromagnetic wave propagating along the positive $z$-axis with an electric field $\textbf{E}=\left(E_x, E_y\right)$ through a 
linear polariser with the axis oriented at an angle $\theta$ with respect to the $x$-direction ($\vu{u}=\left(\cos\theta,\sin\theta\right)$), the power measured after the polariser is $P\left(\theta\right) \propto \abs{\,\mathbf{E} \cdot \vu{u}(\theta) \, }^{2}$. From the measurements of $P\left(\theta\right)$ at each angle of the polariser, while solving an over-determined system of linear equations, one can fit the three parameters \textit{A}, \textit{B} and \textit{C} and subsequently retrieve $\lvert E_x\rvert $, $\lvert E_y\rvert $ and $\cos\left(\psi_y - \psi_x \right)$, where $\psi_y - \psi_x$ represents the phase difference between the $x$ and $y$ components of the field. Since one cannot distinguish between left-handed and right-handed elliptically polarised light, in order to retrieve the exact Jones vector, some assumptions on the global phase and handedness of the polarised light need to be made. The global phase factor between $E_x$ and $E_y$ can be chosen such that $\psi_x=0$, together with the assumption that $\Psi_{y}>0$. These assumptions do not change the shape of the polarisation ellipse. From the fitting parameters \textit{A}, \textit{B} and \textit{C}, one can retrieve the polarisation components at each point of the beam:
\begin{subequations}
\begin{align}
E_x &= \sqrt{A+B} \,,\\
E_y &= \frac{1}{\sqrt{A+B}}\left[C+i\,\sqrt{A^2-B^2-C^2}\right].
\end{align}
\end{subequations}

\subsection{Fitting procedure for modal content analysis}
The modal content of the radial beam is retrieved via a custom-made fitting procedure. This uses the same set of images measured (or calculated) for polarimetry description (Section~\ref{sec:polarimetry}). The measured (calculated) intensity for the linear polariser axis forming an angle $\theta$ with the $x$-axis can be expressed as 
\begin{equation}
  I_{\theta}\left(r, \phi, z\right) = \lvert E_{x}\rvert ^{2} \cos^{2}{\theta} + \lvert E_{y}\rvert ^2 \sin^{2}{\theta} + \Re\left(E_{x} E_{y}^{\ast}\right) \sin 2\theta\,, 
\end{equation}
where $E_x$ and $E_y$ are the components of the unknown electric field reaching the camera. In principle, each component of this field can be written as a superposition of an infinite number of the LG basis modes. At the same time, the number of present modes can be reasonably limited considering that the size of the modes scales up quickly with their quantum numbers ($\ell$ for the azimuthal angle, $p$ for the radial coordinate). The fitting procedure is based on fixing the maximum values for both quantum numbers $(\ell_{\rm max}$ and p$_{\rm max})$) so that a total number of $N = L\cdot \,P$ modes ($L = \ell_{\rm max} + 1$, $P = p_{\rm max} + 1$) is used to decompose the electric field components as follows: 
\begin{align}
E_{x} \left(r, \phi, z\right) &= \sum_{\ell,p = 0}^{\ell_{\rm max},\,p_{\rm max}} A_{\ell p}\, e^{i \varphi^{A}_{\ell p}} \,\, {\rm LG}_{\ell p}\left(r, \phi, z\right)\, \cos\ell \phi\,, \\
E_{y} \left(r, \phi, z\right) &= \sum_{\ell,p = 0}^{\ell_{\rm max},\,p_{\rm max}} B_{\ell p}\, e^{i \varphi^{B}_{\ell p}} \,\, {\rm LG}_{\ell p}(r, \phi, z)\ \sin \ell \phi \,, 
\end{align}\label{eq:e_fit}
where $A_{\ell \rm p},\, B_{\ell \rm p},\, \varphi^A_{\ell \rm p}$ and $\varphi^B_{\ell p}$ are the amplitude and phase of each component, respectively, imposed to be real-valued.
The total number of fitting parameters will be $2(N-1)$, given $\phi^{A_{00}},\ \phi^{B_{00}} = 0$ so that the fitting procedure only considers relative phases. Examples of the intensity maps fed to the fit, the field resulting from the fit and the main modes contained in it are shown in Fig.~\ref{fig:FigS2}.

\subsection{Semi-analytical modelling}
The theoretical study of this system is based on a semi-analytical approach described in detail in Ref.~\cite{aita2024PRB}. The approach is based on the use of the angular spectrum formalism, implemented via Richards-Wolf theory for vectorial diffraction
The developed model can be applied to focused vector beams propagating through multilayered media including layers with anisotropic optical behaviour.

\section{Data Availability}
All the data supporting the findings of this work are presented in the text and available from the corresponding author upon reasonable request.
\newpage
\bibliography{rods-azi-rad.bib} 


\begin{thebibliography}{61}
\ifx \bisbn   \undefined \def \bisbn  #1{ISBN #1}\fi
\ifx \binits  \undefined \def \binits#1{#1}\fi
\ifx \bauthor  \undefined \def \bauthor#1{#1}\fi
\ifx \batitle  \undefined \def \batitle#1{#1}\fi
\ifx \bjtitle  \undefined \def \bjtitle#1{#1}\fi
\ifx \bvolume  \undefined \def \bvolume#1{\textbf{#1}}\fi
\ifx \byear  \undefined \def \byear#1{#1}\fi
\ifx \bissue  \undefined \def \bissue#1{#1}\fi
\ifx \bfpage  \undefined \def \bfpage#1{#1}\fi
\ifx \blpage  \undefined \def \blpage #1{#1}\fi
\ifx \burl  \undefined \def \burl#1{\textsf{#1}}\fi
\ifx \doiurl  \undefined \def \doiurl#1{\url{https://doi.org/#1}}\fi
\ifx \betal  \undefined \def \betal{\textit{et al.}}\fi
\ifx \binstitute  \undefined \def \binstitute#1{#1}\fi
\ifx \binstitutionaled  \undefined \def \binstitutionaled#1{#1}\fi
\ifx \bctitle  \undefined \def \bctitle#1{#1}\fi
\ifx \beditor  \undefined \def \beditor#1{#1}\fi
\ifx \bpublisher  \undefined \def \bpublisher#1{#1}\fi
\ifx \bbtitle  \undefined \def \bbtitle#1{#1}\fi
\ifx \bedition  \undefined \def \bedition#1{#1}\fi
\ifx \bseriesno  \undefined \def \bseriesno#1{#1}\fi
\ifx \blocation  \undefined \def \blocation#1{#1}\fi
\ifx \bsertitle  \undefined \def \bsertitle#1{#1}\fi
\ifx \bsnm \undefined \def \bsnm#1{#1}\fi
\ifx \bsuffix \undefined \def \bsuffix#1{#1}\fi
\ifx \bparticle \undefined \def \bparticle#1{#1}\fi
\ifx \barticle \undefined \def \barticle#1{#1}\fi
\bibcommenthead
\ifx \bconfdate \undefined \def \bconfdate #1{#1}\fi
\ifx \botherref \undefined \def \botherref #1{#1}\fi
\ifx \url \undefined \def \url#1{\textsf{#1}}\fi
\ifx \bchapter \undefined \def \bchapter#1{#1}\fi
\ifx \bbook \undefined \def \bbook#1{#1}\fi
\ifx \bcomment \undefined \def \bcomment#1{#1}\fi
\ifx \oauthor \undefined \def \oauthor#1{#1}\fi
\ifx \citeauthoryear \undefined \def \citeauthoryear#1{#1}\fi
\ifx \endbibitem  \undefined \def \endbibitem {}\fi
\ifx \bconflocation  \undefined \def \bconflocation#1{#1}\fi
\ifx \arxivurl  \undefined \def \arxivurl#1{\textsf{#1}}\fi
\csname PreBibitemsHook\endcsname

\bibitem[\protect\citeauthoryear{Bliokh et~al.}{2023}]{bliokh2023roadmap}
\begin{barticle}
\bauthor{\bsnm{Bliokh}, \binits{K.Y.}},
\bauthor{\bsnm{Karimi}, \binits{E.}},
\bauthor{\bsnm{Padgett}, \binits{M.J.}},
\bauthor{\bsnm{Alonso}, \binits{M.A.}},
\bauthor{\bsnm{Dennis}, \binits{M.R.}},
\bauthor{\bsnm{Dudley}, \binits{A.}},
\bauthor{\bsnm{Forbes}, \binits{A.}},
\bauthor{\bsnm{Zahedpour}, \binits{S.}},
\bauthor{\bsnm{Hancock}, \binits{S.W.}},
\bauthor{\bsnm{Milchberg}, \binits{H.M.}}, \betal:
\batitle{Roadmap on structured waves}.
\bjtitle{Journal of Optics}
\bvolume{25}(\bissue{10}),
\bfpage{103001}
(\byear{2023})
\end{barticle}
\endbibitem

\bibitem[\protect\citeauthoryear{Willner et~al.}{2021}]{Willner2021}
\begin{barticle}
\bauthor{\bsnm{Willner}, \binits{A.E.}},
\bauthor{\bsnm{Pang}, \binits{K.}},
\bauthor{\bsnm{Song}, \binits{H.}},
\bauthor{\bsnm{Zou}, \binits{K.}},
\bauthor{\bsnm{Zhou}, \binits{H.}}:
\batitle{{Orbital angular momentum of light for communications}}.
\bjtitle{Applied Physics Reviews}
\bvolume{8}(\bissue{4}),
\bfpage{041312}
(\byear{2021})
\end{barticle}
\endbibitem

\bibitem[\protect\citeauthoryear{Forbes et~al.}{2021}]{Forbes2021}
\begin{barticle}
\bauthor{\bsnm{Forbes}, \binits{A.}},
\bauthor{\bsnm{Oliveira}, \binits{M.}},
\bauthor{\bsnm{Dennis}, \binits{M.R.}}:
\batitle{Structured light}.
\bjtitle{Nature Photonics}
\bvolume{15}(\bissue{4}),
\bfpage{253}--\blpage{262}
(\byear{2021})
\end{barticle}
\endbibitem

\bibitem[\protect\citeauthoryear{Li et~al.}{2024}]{Li2024}
\begin{barticle}
\bauthor{\bsnm{Li}, \binits{H.}},
\bauthor{\bsnm{Chen}, \binits{F.}},
\bauthor{\bsnm{Jia}, \binits{H.}},
\bauthor{\bsnm{Ye}, \binits{Z.}},
\bauthor{\bsnm{Zhou}, \binits{H.}},
\bauthor{\bsnm{Luo}, \binits{S.}},
\bauthor{\bsnm{Shi}, \binits{J.}},
\bauthor{\bsnm{Sun}, \binits{Z.}},
\bauthor{\bsnm{Xu}, \binits{H.}},
\bauthor{\bsnm{Xu}, \binits{H.}},
\bauthor{\bsnm{Byrnes}, \binits{T.}},
\bauthor{\bsnm{Chen}, \binits{Z.}},
\bauthor{\bsnm{Wu}, \binits{J.}}:
\batitle{All-optical temporal logic gates in localized exciton polaritons}.
\bjtitle{Nature Photonics}
\bvolume{18}(\bissue{8}),
\bfpage{864}--\blpage{869}
(\byear{2024})
\end{barticle}
\endbibitem

\bibitem[\protect\citeauthoryear{Shen et~al.}{2019}]{shen2019optical}
\begin{barticle}
\bauthor{\bsnm{Shen}, \binits{Y.}},
\bauthor{\bsnm{Wang}, \binits{X.}},
\bauthor{\bsnm{Xie}, \binits{Z.}},
\bauthor{\bsnm{Min}, \binits{C.}},
\bauthor{\bsnm{Fu}, \binits{X.}},
\bauthor{\bsnm{Liu}, \binits{Q.}},
\bauthor{\bsnm{Gong}, \binits{M.}},
\bauthor{\bsnm{Yuan}, \binits{X.}}:
\batitle{Optical vortices 30 years on: Oam manipulation from topological charge to multiple singularities}.
\bjtitle{Light: Science \& Applications}
\bvolume{8}(\bissue{1}),
\bfpage{1}--\blpage{29}
(\byear{2019})
\end{barticle}
\endbibitem

\bibitem[\protect\citeauthoryear{Zhan}{2009}]{Zhan09}
\begin{barticle}
\bauthor{\bsnm{Zhan}, \binits{Q.}}:
\batitle{Cylindrical vector beams: from mathematical concepts to applications}.
\bjtitle{Adv. Opt. Photon.}
\bvolume{1}(\bissue{1}),
\bfpage{1}--\blpage{57}
(\byear{2009})
\end{barticle}
\endbibitem

\bibitem[\protect\citeauthoryear{Shen et~al.}{2024}]{shen2024optical}
\begin{barticle}
\bauthor{\bsnm{Shen}, \binits{Y.}},
\bauthor{\bsnm{Zhang}, \binits{Q.}},
\bauthor{\bsnm{Shi}, \binits{P.}},
\bauthor{\bsnm{Du}, \binits{L.}},
\bauthor{\bsnm{Yuan}, \binits{X.}},
\bauthor{\bsnm{Zayats}, \binits{A.V.}}:
\batitle{Optical skyrmions and other topological quasiparticles of light}.
\bjtitle{Nature Photonics}
\bvolume{18}(\bissue{1}),
\bfpage{15}--\blpage{25}
(\byear{2024})
\end{barticle}
\endbibitem

\bibitem[\protect\citeauthoryear{Shen et~al.}{2023}]{shen2023topological}
\begin{barticle}
\bauthor{\bsnm{Shen}, \binits{Y.}},
\bauthor{\bsnm{Yu}, \binits{B.}},
\bauthor{\bsnm{Wu}, \binits{H.}},
\bauthor{\bsnm{Li}, \binits{C.}},
\bauthor{\bsnm{Zhu}, \binits{Z.}},
\bauthor{\bsnm{Zayats}, \binits{A.V.}}:
\batitle{Topological transformation and free-space transport of photonic hopfions}.
\bjtitle{Advanced Photonics}
\bvolume{5}(\bissue{1}),
\bfpage{015001}--\blpage{015001}
(\byear{2023})
\end{barticle}
\endbibitem

\bibitem[\protect\citeauthoryear{Sugic et~al.}{2021}]{sugic2021particle}
\begin{barticle}
\bauthor{\bsnm{Sugic}, \binits{D.}},
\bauthor{\bsnm{Droop}, \binits{R.}},
\bauthor{\bsnm{Otte}, \binits{E.}},
\bauthor{\bsnm{Ehrmanntraut}, \binits{D.}},
\bauthor{\bsnm{Nori}, \binits{F.}},
\bauthor{\bsnm{Ruostekoski}, \binits{J.}},
\bauthor{\bsnm{Denz}, \binits{C.}},
\bauthor{\bsnm{Dennis}, \binits{M.R.}}:
\batitle{Particle-like topologies in light}.
\bjtitle{Nature Communications}
\bvolume{12}(\bissue{1}),
\bfpage{6785}
(\byear{2021})
\end{barticle}
\endbibitem

\bibitem[\protect\citeauthoryear{Singh et~al.}{2023}]{singh2023robust}
\begin{barticle}
\bauthor{\bsnm{Singh}, \binits{K.}},
\bauthor{\bsnm{Nape}, \binits{I.}},
\bauthor{\bsnm{Buono}, \binits{W.T.}},
\bauthor{\bsnm{Dudley}, \binits{A.}},
\bauthor{\bsnm{Forbes}, \binits{A.}}:
\batitle{A robust basis for multi-bit optical communication with vectorial light}.
\bjtitle{Laser \& Photonics Reviews}
\bvolume{17}(\bissue{6}),
\bfpage{2200844}
(\byear{2023})
\end{barticle}
\endbibitem

\bibitem[\protect\citeauthoryear{Aita and Zayats}{2022}]{aita2022enhancement}
\begin{barticle}
\bauthor{\bsnm{Aita}, \binits{V.}},
\bauthor{\bsnm{Zayats}, \binits{A.V.}}:
\batitle{Enhancement of optical spin-orbit coupling in anisotropic enz metamaterials}.
\bjtitle{IEEE Photonics Journal}
\bvolume{15}(\bissue{1}),
\bfpage{1}--\blpage{8}
(\byear{2022})
\end{barticle}
\endbibitem

\bibitem[\protect\citeauthoryear{Afanasev et~al.}{2023}]{Afanasev2023}
\begin{barticle}
\bauthor{\bsnm{Afanasev}, \binits{A.}},
\bauthor{\bsnm{Kingsley-Smith}, \binits{J.}},
\bauthor{\bsnm{Rodr{\'i}guez-Fortu{\~n}o}, \binits{F.J.}},
\bauthor{\bsnm{Zayats}, \binits{A.V.}}:
\batitle{{Nondiffractive three-dimensional polarization features of optical vortex beams}}.
\bjtitle{Advanced Photonics Nexus}
\bvolume{2}(\bissue{2}),
\bfpage{026001}
(\byear{2023})
\end{barticle}
\endbibitem

\bibitem[\protect\citeauthoryear{Aiello et~al.}{2015}]{Aiello2015}
\begin{barticle}
\bauthor{\bsnm{Aiello}, \binits{A.}},
\bauthor{\bsnm{Banzer}, \binits{P.}},
\bauthor{\bsnm{Neugebauer}, \binits{M.}},
\bauthor{\bsnm{Leuchs}, \binits{G.}}:
\batitle{From transverse angular momentum to photonic wheels}.
\bjtitle{Nature Photonics}
\bvolume{9}(\bissue{12}),
\bfpage{789}--\blpage{795}
(\byear{2015})
\end{barticle}
\endbibitem

\bibitem[\protect\citeauthoryear{Aiello and Banzer}{2016}]{Aiello_2016}
\begin{barticle}
\bauthor{\bsnm{Aiello}, \binits{A.}},
\bauthor{\bsnm{Banzer}, \binits{P.}}:
\batitle{The ubiquitous photonic wheel}.
\bjtitle{Journal of Optics}
\bvolume{18}(\bissue{8}),
\bfpage{085605}
(\byear{2016})
\end{barticle}
\endbibitem

\bibitem[\protect\citeauthoryear{Vernon et~al.}{2024}]{Vernon2024}
\begin{barticle}
\bauthor{\bsnm{Vernon}, \binits{A.J.}},
\bauthor{\bsnm{Golat}, \binits{S.}},
\bauthor{\bsnm{Rigouzzo}, \binits{C.}},
\bauthor{\bsnm{Lim}, \binits{E.A.}},
\bauthor{\bsnm{Rodr{\'i}guez-Fortu{\~{n}}o}, \binits{F.J.}}:
\batitle{A decomposition of light's spin angular momentum density}.
\bjtitle{Light: Science {\&} Applications}
\bvolume{13}(\bissue{1}),
\bfpage{160}
(\byear{2024})
\end{barticle}
\endbibitem

\bibitem[\protect\citeauthoryear{Eismann et~al.}{2021}]{eismann2021transverse}
\begin{barticle}
\bauthor{\bsnm{Eismann}, \binits{J.}},
\bauthor{\bsnm{Nicholls}, \binits{L.}},
\bauthor{\bsnm{Roth}, \binits{D.}},
\bauthor{\bsnm{Alonso}, \binits{M.A.}},
\bauthor{\bsnm{Banzer}, \binits{P.}},
\bauthor{\bsnm{Rodr{\'\i}guez-Fortu{\~n}o}, \binits{F.}},
\bauthor{\bsnm{Zayats}, \binits{A.}},
\bauthor{\bsnm{Nori}, \binits{F.}},
\bauthor{\bsnm{Bliokh}, \binits{K.}}:
\batitle{Transverse spinning of unpolarized light}.
\bjtitle{Nature Photonics}
\bvolume{15}(\bissue{2}),
\bfpage{156}--\blpage{161}
(\byear{2021})
\end{barticle}
\endbibitem

\bibitem[\protect\citeauthoryear{Luxmoore et~al.}{2013}]{Luxmoore2013}
\begin{barticle}
\bauthor{\bsnm{Luxmoore}, \binits{I.J.}},
\bauthor{\bsnm{Wasley}, \binits{N.A.}},
\bauthor{\bsnm{Ramsay}, \binits{A.J.}},
\bauthor{\bsnm{Thijssen}, \binits{A.C.T.}},
\bauthor{\bsnm{Oulton}, \binits{R.}},
\bauthor{\bsnm{Hugues}, \binits{M.}},
\bauthor{\bsnm{Kasture}, \binits{S.}},
\bauthor{\bsnm{Achanta}, \binits{V.G.}},
\bauthor{\bsnm{Fox}, \binits{A.M.}},
\bauthor{\bsnm{Skolnick}, \binits{M.S.}}:
\batitle{Interfacing spins in an ingaas quantum dot to a semiconductor waveguide circuit using emitted photons}.
\bjtitle{Phys. Rev. Lett.}
\bvolume{110},
\bfpage{037402}
(\byear{2013})
\end{barticle}
\endbibitem

\bibitem[\protect\citeauthoryear{Junge et~al.}{2013}]{Junge2013}
\begin{barticle}
\bauthor{\bsnm{Junge}, \binits{C.}},
\bauthor{\bsnm{O'Shea}, \binits{D.}},
\bauthor{\bsnm{Volz}, \binits{J.}},
\bauthor{\bsnm{Rauschenbeutel}, \binits{A.}}:
\batitle{Strong coupling between single atoms and nontransversal photons}.
\bjtitle{Phys. Rev. Lett.}
\bvolume{110},
\bfpage{213604}
(\byear{2013})
\end{barticle}
\endbibitem

\bibitem[\protect\citeauthoryear{Rodríguez-Fortuño et~al.}{2013}]{Fortuño2013}
\begin{barticle}
\bauthor{\bsnm{Rodríguez-Fortuño}, \binits{F.J.}},
\bauthor{\bsnm{Marino}, \binits{G.}},
\bauthor{\bsnm{Ginzburg}, \binits{P.}},
\bauthor{\bsnm{O’Connor}, \binits{D.}},
\bauthor{\bsnm{Martínez}, \binits{A.}},
\bauthor{\bsnm{Wurtz}, \binits{G.A.}},
\bauthor{\bsnm{Zayats}, \binits{A.V.}}:
\batitle{Near-field interference for the unidirectional excitation of electromagnetic guided modes}.
\bjtitle{Science}
\bvolume{340}(\bissue{6130}),
\bfpage{328}--\blpage{330}
(\byear{2013})
\end{barticle}
\endbibitem

\bibitem[\protect\citeauthoryear{Petersen et~al.}{2014}]{Petersen2014}
\begin{barticle}
\bauthor{\bsnm{Petersen}, \binits{J.}},
\bauthor{\bsnm{Volz}, \binits{J.}},
\bauthor{\bsnm{Rauschenbeutel}, \binits{A.}}:
\batitle{Chiral nanophotonic waveguide interface based on spin-orbit interaction of light}.
\bjtitle{Science}
\bvolume{346}(\bissue{6205}),
\bfpage{67}--\blpage{71}
(\byear{2014})
\end{barticle}
\endbibitem

\bibitem[\protect\citeauthoryear{Bliokh et~al.}{2015}]{Bliokh2015}
\begin{barticle}
\bauthor{\bsnm{Bliokh}, \binits{K.Y.}},
\bauthor{\bsnm{Rodr{\'i}guez-Fortu{\~{n}}o}, \binits{F.J.}},
\bauthor{\bsnm{Nori}, \binits{F.}},
\bauthor{\bsnm{Zayats}, \binits{A.V.}}:
\batitle{Spin--orbit interactions of light}.
\bjtitle{Nature Photonics}
\bvolume{9}(\bissue{12}),
\bfpage{796}--\blpage{808}
(\byear{2015})
\end{barticle}
\endbibitem

\bibitem[\protect\citeauthoryear{Rodr{\'i}guez-Fortu{\~{n}}o et~al.}{2015}]{Rodríguez-Fortuño2015}
\begin{barticle}
\bauthor{\bsnm{Rodr{\'i}guez-Fortu{\~{n}}o}, \binits{F.J.}},
\bauthor{\bsnm{Engheta}, \binits{N.}},
\bauthor{\bsnm{Mart{\'i}nez}, \binits{A.}},
\bauthor{\bsnm{Zayats}, \binits{A.V.}}:
\batitle{Lateral forces on circularly polarizable particles near a surface}.
\bjtitle{Nature Communications}
\bvolume{6}(\bissue{1}),
\bfpage{8799}
(\byear{2015})
\end{barticle}
\endbibitem

\bibitem[\protect\citeauthoryear{Kapitanova et~al.}{2014}]{Kapitanova2014}
\begin{barticle}
\bauthor{\bsnm{Kapitanova}, \binits{P.V.}},
\bauthor{\bsnm{Ginzburg}, \binits{P.}},
\bauthor{\bsnm{Rodr{\'i}guez-Fortu{\~{n}}o}, \binits{F.J.}},
\bauthor{\bsnm{Filonov}, \binits{D.S.}},
\bauthor{\bsnm{Voroshilov}, \binits{P.M.}},
\bauthor{\bsnm{Belov}, \binits{P.A.}},
\bauthor{\bsnm{Poddubny}, \binits{A.N.}},
\bauthor{\bsnm{Kivshar}, \binits{Y.S.}},
\bauthor{\bsnm{Wurtz}, \binits{G.A.}},
\bauthor{\bsnm{Zayats}, \binits{A.V.}}:
\batitle{Photonic spin hall effect in hyperbolic metamaterials for polarization-controlled routing of subwavelength modes}.
\bjtitle{Nature Communications}
\bvolume{5}(\bissue{1}),
\bfpage{3226}
(\byear{2014})
\end{barticle}
\endbibitem

\bibitem[\protect\citeauthoryear{le~Feber et~al.}{2015}]{leFeber2015}
\begin{barticle}
\bauthor{\bsnm{Feber}, \binits{B.}},
\bauthor{\bsnm{Rotenberg}, \binits{N.}},
\bauthor{\bsnm{Kuipers}, \binits{L.}}:
\batitle{Nanophotonic control of circular dipole emission}.
\bjtitle{Nature Communications}
\bvolume{6}(\bissue{1}),
\bfpage{6695}
(\byear{2015})
\end{barticle}
\endbibitem

\bibitem[\protect\citeauthoryear{Neugebauer et~al.}{2015}]{Neugebauer2015}
\begin{barticle}
\bauthor{\bsnm{Neugebauer}, \binits{M.}},
\bauthor{\bsnm{Bauer}, \binits{T.}},
\bauthor{\bsnm{Aiello}, \binits{A.}},
\bauthor{\bsnm{Banzer}, \binits{P.}}:
\batitle{Measuring the transverse spin density of light}.
\bjtitle{Phys. Rev. Lett.}
\bvolume{114},
\bfpage{063901}
(\byear{2015})
\end{barticle}
\endbibitem

\bibitem[\protect\citeauthoryear{F.~Picardi et~al.}{2019}]{Picardi2019}
\begin{barticle}
\bauthor{\bsnm{F.~Picardi}, \binits{M.}},
\bauthor{\bsnm{V.~Zayats}, \binits{A.}},
\bauthor{\bsnm{J.~Rodríguez-Fortuño}, \binits{F.}}:
\batitle{Amplitude and phase control of guided modes excitation from a single dipole source: Engineering far- and near-field directionality}.
\bjtitle{Laser \& Photonics Reviews}
\bvolume{13}(\bissue{12}),
\bfpage{1900250}
(\byear{2019})
\end{barticle}
\endbibitem

\bibitem[\protect\citeauthoryear{Krasavin et~al.}{2018}]{krasavin2018generalization}
\begin{barticle}
\bauthor{\bsnm{Krasavin}, \binits{A.V.}},
\bauthor{\bsnm{Segovia}, \binits{P.}},
\bauthor{\bsnm{Dubrovka}, \binits{R.}},
\bauthor{\bsnm{Olivier}, \binits{N.}},
\bauthor{\bsnm{Wurtz}, \binits{G.A.}},
\bauthor{\bsnm{Ginzburg}, \binits{P.}},
\bauthor{\bsnm{Zayats}, \binits{A.V.}}:
\batitle{Generalization of the optical theorem: experimental proof for radially polarized beams}.
\bjtitle{Light: Science \& Applications}
\bvolume{7}(\bissue{1}),
\bfpage{1}--\blpage{10}
(\byear{2018})
\end{barticle}
\endbibitem

\bibitem[\protect\citeauthoryear{Naidoo et~al.}{2016}]{naidoo2016controlled}
\begin{barticle}
\bauthor{\bsnm{Naidoo}, \binits{D.}},
\bauthor{\bsnm{Roux}, \binits{F.S.}},
\bauthor{\bsnm{Dudley}, \binits{A.}},
\bauthor{\bsnm{Litvin}, \binits{I.}},
\bauthor{\bsnm{Piccirillo}, \binits{B.}},
\bauthor{\bsnm{Marrucci}, \binits{L.}},
\bauthor{\bsnm{Forbes}, \binits{A.}}:
\batitle{Controlled generation of higher-order poincar{\'e} sphere beams from a laser}.
\bjtitle{Nature Photonics}
\bvolume{10}(\bissue{5}),
\bfpage{327}--\blpage{332}
(\byear{2016})
\end{barticle}
\endbibitem

\bibitem[\protect\citeauthoryear{Rosales-Guzm{\'a}n et~al.}{2018}]{rosales2018review}
\begin{barticle}
\bauthor{\bsnm{Rosales-Guzm{\'a}n}, \binits{C.}},
\bauthor{\bsnm{Ndagano}, \binits{B.}},
\bauthor{\bsnm{Forbes}, \binits{A.}}:
\batitle{A review of complex vector light fields and their applications}.
\bjtitle{Journal of Optics}
\bvolume{20}(\bissue{12}),
\bfpage{123001}
(\byear{2018})
\end{barticle}
\endbibitem

\bibitem[\protect\citeauthoryear{Karimi et~al.}{2010}]{karimi2010polarization}
\begin{barticle}
\bauthor{\bsnm{Karimi}, \binits{E.}},
\bauthor{\bsnm{Slussarenko}, \binits{S.}},
\bauthor{\bsnm{Piccirillo}, \binits{B.}},
\bauthor{\bsnm{Marrucci}, \binits{L.}},
\bauthor{\bsnm{Santamato}, \binits{E.}}:
\batitle{Polarization-controlled evolution of light transverse modes and associated pancharatnam geometric phase in orbital angular momentum}.
\bjtitle{Physical Review A}
\bvolume{81}(\bissue{5}),
\bfpage{053813}
(\byear{2010})
\end{barticle}
\endbibitem

\bibitem[\protect\citeauthoryear{Liu et~al.}{2017}]{liu2017generation}
\begin{barticle}
\bauthor{\bsnm{Liu}, \binits{Z.}},
\bauthor{\bsnm{Liu}, \binits{Y.}},
\bauthor{\bsnm{Ke}, \binits{Y.}},
\bauthor{\bsnm{Liu}, \binits{Y.}},
\bauthor{\bsnm{Shu}, \binits{W.}},
\bauthor{\bsnm{Luo}, \binits{H.}},
\bauthor{\bsnm{Wen}, \binits{S.}}:
\batitle{Generation of arbitrary vector vortex beams on hybrid-order poincar{\'e} sphere}.
\bjtitle{Photonics Research}
\bvolume{5}(\bissue{1}),
\bfpage{15}--\blpage{21}
(\byear{2017})
\end{barticle}
\endbibitem

\bibitem[\protect\citeauthoryear{Brasselet et~al.}{2009}]{brasselet2009optical}
\begin{barticle}
\bauthor{\bsnm{Brasselet}, \binits{E.}},
\bauthor{\bsnm{Murazawa}, \binits{N.}},
\bauthor{\bsnm{Misawa}, \binits{H.}},
\bauthor{\bsnm{Juodkazis}, \binits{S.}}:
\batitle{Optical vortices from liquid crystal droplets}.
\bjtitle{Physical Review Letters}
\bvolume{103}(\bissue{10}),
\bfpage{103903}
(\byear{2009})
\end{barticle}
\endbibitem

\bibitem[\protect\citeauthoryear{Rumala et~al.}{2013}]{rumala2013tunable}
\begin{barticle}
\bauthor{\bsnm{Rumala}, \binits{Y.S.}},
\bauthor{\bsnm{Milione}, \binits{G.}},
\bauthor{\bsnm{Nguyen}, \binits{T.A.}},
\bauthor{\bsnm{Pratavieira}, \binits{S.}},
\bauthor{\bsnm{Hossain}, \binits{Z.}},
\bauthor{\bsnm{Nolan}, \binits{D.}},
\bauthor{\bsnm{Slussarenko}, \binits{S.}},
\bauthor{\bsnm{Karimi}, \binits{E.}},
\bauthor{\bsnm{Marrucci}, \binits{L.}},
\bauthor{\bsnm{Alfano}, \binits{R.R.}}:
\batitle{Tunable supercontinuum light vector vortex beam generator using a q-plate}.
\bjtitle{Optics letters}
\bvolume{38}(\bissue{23}),
\bfpage{5083}--\blpage{5086}
(\byear{2013})
\end{barticle}
\endbibitem

\bibitem[\protect\citeauthoryear{Chen et~al.}{2015}]{chen2015generation}
\begin{barticle}
\bauthor{\bsnm{Chen}, \binits{P.}},
\bauthor{\bsnm{Ji}, \binits{W.}},
\bauthor{\bsnm{Wei}, \binits{B.-Y.}},
\bauthor{\bsnm{Hu}, \binits{W.}},
\bauthor{\bsnm{Chigrinov}, \binits{V.}},
\bauthor{\bsnm{Lu}, \binits{Y.-Q.}}:
\batitle{Generation of arbitrary vector beams with liquid crystal polarization converters and vector-photoaligned q-plates}.
\bjtitle{Applied Physics Letters}
\bvolume{107}(\bissue{24}),
\bfpage{241102}
(\byear{2015})
\end{barticle}
\endbibitem

\bibitem[\protect\citeauthoryear{Liu et~al.}{2018}]{liu2018highly}
\begin{barticle}
\bauthor{\bsnm{Liu}, \binits{S.}},
\bauthor{\bsnm{Qi}, \binits{S.}},
\bauthor{\bsnm{Zhang}, \binits{Y.}},
\bauthor{\bsnm{Li}, \binits{P.}},
\bauthor{\bsnm{Wu}, \binits{D.}},
\bauthor{\bsnm{Han}, \binits{L.}},
\bauthor{\bsnm{Zhao}, \binits{J.}}:
\batitle{Highly efficient generation of arbitrary vector beams with tunable polarization, phase, and amplitude}.
\bjtitle{Photonics Research}
\bvolume{6}(\bissue{4}),
\bfpage{228}--\blpage{233}
(\byear{2018})
\end{barticle}
\endbibitem

\bibitem[\protect\citeauthoryear{Perez-Garcia et~al.}{2017}]{perez2017demand}
\begin{barticle}
\bauthor{\bsnm{Perez-Garcia}, \binits{B.}},
\bauthor{\bsnm{L{\'o}pez-Mariscal}, \binits{C.}},
\bauthor{\bsnm{Hernandez-Aranda}, \binits{R.I.}},
\bauthor{\bsnm{Guti{\'e}rrez-Vega}, \binits{J.C.}}:
\batitle{On-demand tailored vector beams}.
\bjtitle{Applied optics}
\bvolume{56}(\bissue{24}),
\bfpage{6967}--\blpage{6972}
(\byear{2017})
\end{barticle}
\endbibitem

\bibitem[\protect\citeauthoryear{Radwell et~al.}{2016}]{radwell2016achromatic}
\begin{barticle}
\bauthor{\bsnm{Radwell}, \binits{N.}},
\bauthor{\bsnm{Hawley}, \binits{R.}},
\bauthor{\bsnm{G{\"o}tte}, \binits{J.}},
\bauthor{\bsnm{Franke-Arnold}, \binits{S.}}:
\batitle{Achromatic vector vortex beams from a glass cone}.
\bjtitle{Nature Communications}
\bvolume{7}(\bissue{1}),
\bfpage{10564}
(\byear{2016})
\end{barticle}
\endbibitem

\bibitem[\protect\citeauthoryear{Yue et~al.}{2016}]{yue2016vector}
\begin{barticle}
\bauthor{\bsnm{Yue}, \binits{F.}},
\bauthor{\bsnm{Wen}, \binits{D.}},
\bauthor{\bsnm{Xin}, \binits{J.}},
\bauthor{\bsnm{Gerardot}, \binits{B.D.}},
\bauthor{\bsnm{Li}, \binits{J.}},
\bauthor{\bsnm{Chen}, \binits{X.}}:
\batitle{Vector vortex beam generation with a single plasmonic metasurface}.
\bjtitle{ACS Photonics}
\bvolume{3}(\bissue{9}),
\bfpage{1558}--\blpage{1563}
(\byear{2016})
\end{barticle}
\endbibitem

\bibitem[\protect\citeauthoryear{Chen et~al.}{2016}]{chen2016geometric}
\begin{barticle}
\bauthor{\bsnm{Chen}, \binits{S.}},
\bauthor{\bsnm{Cai}, \binits{Y.}},
\bauthor{\bsnm{Li}, \binits{G.}},
\bauthor{\bsnm{Zhang}, \binits{S.}},
\bauthor{\bsnm{Cheah}, \binits{K.W.}}:
\batitle{Geometric metasurface fork gratings for vortex-beam generation and manipulation}.
\bjtitle{Laser \& Photonics Reviews}
\bvolume{10}(\bissue{2}),
\bfpage{322}--\blpage{326}
(\byear{2016})
\end{barticle}
\endbibitem

\bibitem[\protect\citeauthoryear{Ding et~al.}{2020}]{ding2020focused}
\begin{barticle}
\bauthor{\bsnm{Ding}, \binits{F.}},
\bauthor{\bsnm{Chen}, \binits{Y.}},
\bauthor{\bsnm{Bozhevolnyi}, \binits{S.I.}}:
\batitle{Focused vortex-beam generation using gap-surface plasmon metasurfaces}.
\bjtitle{Nanophotonics}
\bvolume{9}(\bissue{2}),
\bfpage{371}--\blpage{378}
(\byear{2020})
\end{barticle}
\endbibitem

\bibitem[\protect\citeauthoryear{Crameri et~al.}{2020}]{crameri2020misuse}
\begin{barticle}
\bauthor{\bsnm{Crameri}, \binits{F.}},
\bauthor{\bsnm{Shephard}, \binits{G.E.}},
\bauthor{\bsnm{Heron}, \binits{P.J.}}:
\batitle{The misuse of colour in science communication}.
\bjtitle{Nature Communications}
\bvolume{11}(\bissue{1}),
\bfpage{5444}
(\byear{2020})
\end{barticle}
\endbibitem

\bibitem[\protect\citeauthoryear{Gori et~al.}{1987}]{gori1987bessel}
\begin{barticle}
\bauthor{\bsnm{Gori}, \binits{F.}},
\bauthor{\bsnm{Guattari}, \binits{G.}},
\bauthor{\bsnm{Padovani}, \binits{C.}}:
\batitle{Bessel-gauss beams}.
\bjtitle{Optics Communications}
\bvolume{64}(\bissue{6}),
\bfpage{491}--\blpage{495}
(\byear{1987})
\end{barticle}
\endbibitem

\bibitem[\protect\citeauthoryear{Maurer et~al.}{2007}]{maurer2007tailoring}
\begin{barticle}
\bauthor{\bsnm{Maurer}, \binits{C.}},
\bauthor{\bsnm{Jesacher}, \binits{A.}},
\bauthor{\bsnm{F{\"u}rhapter}, \binits{S.}},
\bauthor{\bsnm{Bernet}, \binits{S.}},
\bauthor{\bsnm{Ritsch-Marte}, \binits{M.}}:
\batitle{Tailoring of arbitrary optical vector beams}.
\bjtitle{New Journal off Physics}
\bvolume{9}(\bissue{3}),
\bfpage{78}
(\byear{2007})
\end{barticle}
\endbibitem

\bibitem[\protect\citeauthoryear{Moreno et~al.}{2010}]{moreno2010decomposition}
\begin{barticle}
\bauthor{\bsnm{Moreno}, \binits{I.}},
\bauthor{\bsnm{Davis}, \binits{J.A.}},
\bauthor{\bsnm{Ruiz}, \binits{I.}},
\bauthor{\bsnm{Cottrell}, \binits{D.M.}}:
\batitle{Decomposition of radially and azimuthally polarized beams using a circular-polarization and vortex-sensing diffraction grating}.
\bjtitle{Optics Express}
\bvolume{18}(\bissue{7}),
\bfpage{7173}--\blpage{7183}
(\byear{2010})
\end{barticle}
\endbibitem

\bibitem[\protect\citeauthoryear{Kotlyar et~al.}{2022}]{kotlyar2022index}
\begin{barticle}
\bauthor{\bsnm{Kotlyar}, \binits{V.}},
\bauthor{\bsnm{Kovalev}, \binits{A.}},
\bauthor{\bsnm{Stafeev}, \binits{S.}},
\bauthor{\bsnm{Zaitsev}, \binits{V.}}:
\batitle{Index of the polarization singularity of poincare beams}.
\bjtitle{Bulletin of the Russian Academy of Sciences: Physics}
\bvolume{86}(\bissue{10}),
\bfpage{1158}--\blpage{1163}
(\byear{2022})
\end{barticle}
\endbibitem

\bibitem[\protect\citeauthoryear{Forbes et~al.}{2021}]{forbes2021relevance}
\begin{barticle}
\bauthor{\bsnm{Forbes}, \binits{K.A.}},
\bauthor{\bsnm{Green}, \binits{D.}},
\bauthor{\bsnm{Jones}, \binits{G.A.}}:
\batitle{Relevance of longitudinal fields of paraxial optical vortices}.
\bjtitle{Journal of Optics}
\bvolume{23}(\bissue{7}),
\bfpage{075401}
(\byear{2021})
\end{barticle}
\endbibitem

\bibitem[\protect\citeauthoryear{Roth et~al.}{2024}]{Roth2024}
\begin{barticle}
\bauthor{\bsnm{Roth}, \binits{D.J.}},
\bauthor{\bsnm{Krasavin}, \binits{A.V.}},
\bauthor{\bsnm{Zayats}, \binits{A.V.}}:
\batitle{Nanophotonics with plasmonic nanorod metamaterials}.
\bjtitle{Laser \& Photonics Reviews}
\bvolume{18}(\bissue{8}),
\bfpage{2300886}
(\byear{2024})
\end{barticle}
\endbibitem

\bibitem[\protect\citeauthoryear{Elser et~al.}{2006}]{elser2006nanowire}
\begin{barticle}
\bauthor{\bsnm{Elser}, \binits{J.}},
\bauthor{\bsnm{Wangberg}, \binits{R.}},
\bauthor{\bsnm{Podolskiy}, \binits{V.A.}},
\bauthor{\bsnm{Narimanov}, \binits{E.E.}}:
\batitle{Nanowire metamaterials with extreme optical anisotropy}.
\bjtitle{Applied Physics Letters}
\bvolume{89}(\bissue{26}),
\bfpage{261102}
(\byear{2006})
\end{barticle}
\endbibitem

\bibitem[\protect\citeauthoryear{Johnson and Christy}{1972}]{johnson1972optical}
\begin{barticle}
\bauthor{\bsnm{Johnson}, \binits{P.B.}},
\bauthor{\bsnm{Christy}, \binits{R.-W.}}:
\batitle{Optical constants of the noble metals}.
\bjtitle{Physical Review B}
\bvolume{6}(\bissue{12}),
\bfpage{4370}
(\byear{1972})
\end{barticle}
\endbibitem

\bibitem[\protect\citeauthoryear{Etchegoin et~al.}{2006}]{etchegoin2006analytic}
\begin{barticle}
\bauthor{\bsnm{Etchegoin}, \binits{P.G.}},
\bauthor{\bsnm{Le~Ru}, \binits{E.}},
\bauthor{\bsnm{Meyer}, \binits{M.}}:
\batitle{An analytic model for the optical properties of gold}.
\bjtitle{The Journal of Chemical Physics}
\bvolume{125}(\bissue{16}),
\bfpage{164705}
(\byear{2006})
\end{barticle}
\endbibitem

\bibitem[\protect\citeauthoryear{Etchegoin et~al.}{}]{etchegoinerratum}
\begin{botherref}
\oauthor{\bsnm{Etchegoin}, \binits{P.}},
\oauthor{\bsnm{Le~Ru}, \binits{E.}},
\oauthor{\bsnm{Meyer}, \binits{M.}}:
Erratum:“an analytic model for the optical properties of gold”[j. chem. phys.,()].
The Journal of Chemical Physics
\end{botherref}
\endbibitem

\bibitem[\protect\citeauthoryear{Aita et~al.}{2024}]{aita2024PRB}
\begin{barticle}
\bauthor{\bsnm{Aita}, \binits{V.}},
\bauthor{\bsnm{Shevchenko}, \binits{M.}},
\bauthor{\bsnm{Rodr\'{\i}guez-Fortu\~no}, \binits{F.J.}},
\bauthor{\bsnm{Zayats}, \binits{A.V.}}:
\batitle{Propagation of focused scalar and vector vortex beams in anisotropic media: A semianalytical approach}.
\bjtitle{Physical Review B}
\bvolume{109},
\bfpage{125433}
(\byear{2024})
\end{barticle}
\endbibitem

\bibitem[\protect\citeauthoryear{Ginzburg et~al.}{2013}]{Ginzburg2013}
\begin{barticle}
\bauthor{\bsnm{Ginzburg}, \binits{P.}},
\bauthor{\bsnm{{n}o}, \binits{F.J.R.F.}},
\bauthor{\bsnm{Wurtz}, \binits{G.A.}},
\bauthor{\bsnm{Dickson}, \binits{W.}},
\bauthor{\bsnm{Murphy}, \binits{A.}},
\bauthor{\bsnm{Morgan}, \binits{F.}},
\bauthor{\bsnm{Pollard}, \binits{R.J.}},
\bauthor{\bsnm{Iorsh}, \binits{I.}},
\bauthor{\bsnm{Atrashchenko}, \binits{A.}},
\bauthor{\bsnm{Belov}, \binits{P.A.}},
\bauthor{\bsnm{Kivshar}, \binits{Y.S.}},
\bauthor{\bsnm{Nevet}, \binits{A.}},
\bauthor{\bsnm{Ankonina}, \binits{G.}},
\bauthor{\bsnm{Orenstein}, \binits{M.}},
\bauthor{\bsnm{Zayats}, \binits{A.V.}}:
\batitle{Manipulating polarization of light with ultrathin epsilon-near-zero metamaterials}.
\bjtitle{Opt. Express}
\bvolume{21}(\bissue{12}),
\bfpage{14907}--\blpage{14917}
(\byear{2013})
\end{barticle}
\endbibitem

\bibitem[\protect\citeauthoryear{Evans et~al.}{2006}]{evans2006growth}
\begin{barticle}
\bauthor{\bsnm{Evans}, \binits{P.}},
\bauthor{\bsnm{Hendren}, \binits{W.}},
\bauthor{\bsnm{Atkinson}, \binits{R.}},
\bauthor{\bsnm{Wurtz}, \binits{G.}},
\bauthor{\bsnm{Dickson}, \binits{W.}},
\bauthor{\bsnm{Zayats}, \binits{A.}},
\bauthor{\bsnm{Pollard}, \binits{R.}}:
\batitle{Growth and properties of gold and nickel nanorods in thin film alumina}.
\bjtitle{Nanotechnology}
\bvolume{17}(\bissue{23}),
\bfpage{5746}
(\byear{2006})
\end{barticle}
\endbibitem

\bibitem[\protect\citeauthoryear{Zaleska et~al.}{2024}]{Zaleska2024}
\begin{barticle}
\bauthor{\bsnm{Zaleska}, \binits{A.}},
\bauthor{\bsnm{Krasavin}, \binits{A.V.}},
\bauthor{\bsnm{Zayats}, \binits{A.V.}},
\bauthor{\bsnm{Dickson}, \binits{W.}}:
\batitle{Copper-based core–shell metamaterials with ultra-broadband and reversible enz tunability}.
\bjtitle{Mater. Adv.}
\bvolume{5},
\bfpage{5845}--\blpage{5854}
(\byear{2024})
\end{barticle}
\endbibitem

\bibitem[\protect\citeauthoryear{Dodge}{1986}]{dodge1986alumina}
\begin{bchapter}
\bauthor{\bsnm{Dodge}, \binits{M.J.}}:
\bctitle{Refractive index}.
In: \beditor{\bsnm{Weber}, \binits{M.J.}} (ed.)
\bbtitle{Handbook of Laser Science and Technology, Volume IV: Optical Materials, Part 2}.
\bpublisher{CRC Press}, \blocation{???}
(\byear{1986})
\end{bchapter}
\endbibitem

\bibitem[\protect\citeauthoryear{Lissberger and Nelson}{1974}]{lissberger1974optical}
\begin{barticle}
\bauthor{\bsnm{Lissberger}, \binits{P.}},
\bauthor{\bsnm{Nelson}, \binits{R.}}:
\batitle{Optical properties of thin film au-mgf2 cermets}.
\bjtitle{Thin Solid Films}
\bvolume{21}(\bissue{1}),
\bfpage{159}--\blpage{172}
(\byear{1974})
\end{barticle}
\endbibitem

\bibitem[\protect\citeauthoryear{Born et~al.}{1999}]{Born_Wolf1999}
\begin{bbook}
\bauthor{\bsnm{Born}, \binits{M.}},
\bauthor{\bsnm{Wolf}, \binits{E.}},
\bauthor{\bsnm{Bhatia}, \binits{A.B.}},
\bauthor{\bsnm{Clemmow}, \binits{P.C.}},
\bauthor{\bsnm{Gabor}, \binits{D.}},
\bauthor{\bsnm{Stokes}, \binits{A.R.}},
\bauthor{\bsnm{Taylor}, \binits{A.M.}},
\bauthor{\bsnm{Wayman}, \binits{P.A.}},
\bauthor{\bsnm{Wilcock}, \binits{W.L.}}:
\bbtitle{Principles of Optics: Electromagnetic Theory of Propagation, Interference and Diffraction of Light},
\bedition{7}th edn.
\bpublisher{Cambridge University Press}, \blocation{???}
(\byear{1999})
\end{bbook}
\endbibitem

\bibitem[\protect\citeauthoryear{Fu et~al.}{2016}]{Fu2016}
\begin{barticle}
\bauthor{\bsnm{Fu}, \binits{S.}},
\bauthor{\bsnm{Gao}, \binits{C.}},
\bauthor{\bsnm{Wang}, \binits{T.}},
\bauthor{\bsnm{Zhang}, \binits{S.}},
\bauthor{\bsnm{Zhai}, \binits{Y.}}:
\batitle{{Simultaneous generation of multiple perfect polarization vortices with selective spatial states in various diffraction orders}}.
\bjtitle{Optics Letters}
\bvolume{41}(\bissue{23}),
\bfpage{5454}
(\byear{2016})
\end{barticle}
\endbibitem

\bibitem[\protect\citeauthoryear{Wang et~al.}{2019}]{Wang2019}
\begin{barticle}
\bauthor{\bsnm{Wang}, \binits{P.}},
\bauthor{\bsnm{Liu}, \binits{J.}},
\bauthor{\bsnm{He}, \binits{Y.}},
\bauthor{\bsnm{Zhou}, \binits{X.}},
\bauthor{\bsnm{Ye}, \binits{H.}},
\bauthor{\bsnm{Li}, \binits{Y.}},
\bauthor{\bsnm{Chen}, \binits{S.}},
\bauthor{\bsnm{Fan}, \binits{D.}}:
\batitle{{Arbitrary Cylindrical Vector Beam Generation Using Cross-Polarized Modulation}}.
\bjtitle{IEEE Photonics Technology Letters}
\bvolume{31}(\bissue{11}),
\bfpage{873}--\blpage{876}
(\byear{2019})
\end{barticle}
\endbibitem

\bibitem[\protect\citeauthoryear{Rosales-Guzm{\'a}n and Forbes}{2017}]{rosales2017shape}
\begin{bchapter}
\bauthor{\bsnm{Rosales-Guzm{\'a}n}, \binits{C.}},
\bauthor{\bsnm{Forbes}, \binits{A.}}:
\bctitle{How to shape light with spatial light modulators}.
(\byear{2017}).
\bcomment{Society of Photo-Optical Instrumentation Engineers (SPIE)}
\end{bchapter}
\endbibitem

\end{thebibliography}

\newpage 
\clearpage

\bmhead{Acknowledgements}
This work was supported in part the ERC iCOMM project (789340) and the UK EPSRC project EP/Y015673/1. The authors thank Ryo Mizuta Graphics for providing open-access illustration material for optical components, some of them used in Fig.~\ref{fig:Figure6}. 

\bmhead{Author contributions}
V.A. and D.J.R. contributed equally to this work. V.A., D.J.R. and L.H.N. performed optical measurements; A.Z. fabricated the samples; A.V.K. performed numerical simulations; V.A., M.S. and F.J.R.F. developed the semi-analytical model. A.V.Z. developed the idea and supervised the project. 
All authors contributed to writing the manuscript.

\bmhead{Competing interests}
The authors declare no competing interests

\bmhead{Supplementary information}
Supplementary information for this article is available. 

\newpage 
\begin{center}
\large{\bf Supporting Information\\  Longitudinal field controls vector vortex beams in anisotropic epsilon-near-zero metamaterials}\\
\medskip
\large{Vittorio Aita, Diane J. Roth, Anastasia Zaleska, Alexey V. Krasavin, Luke H. Nicholls, Mykyta Shevchenko, Francisco J. Rodríguez-Fortuño, and Anatoly V. Zayats}\\

\end{center}
\renewcommand{\thepage}{S\arabic{page}} 
\setcounter{page}{1}
\renewcommand{\figurename}{Supplementary Figure}
\setcounter{figure}{0}  
\renewcommand{\thesection}{S\arabic{section}}  
\setcounter{section}{0}


\begin{figure}[h]
    \centering
    \includegraphics[width =  \linewidth]{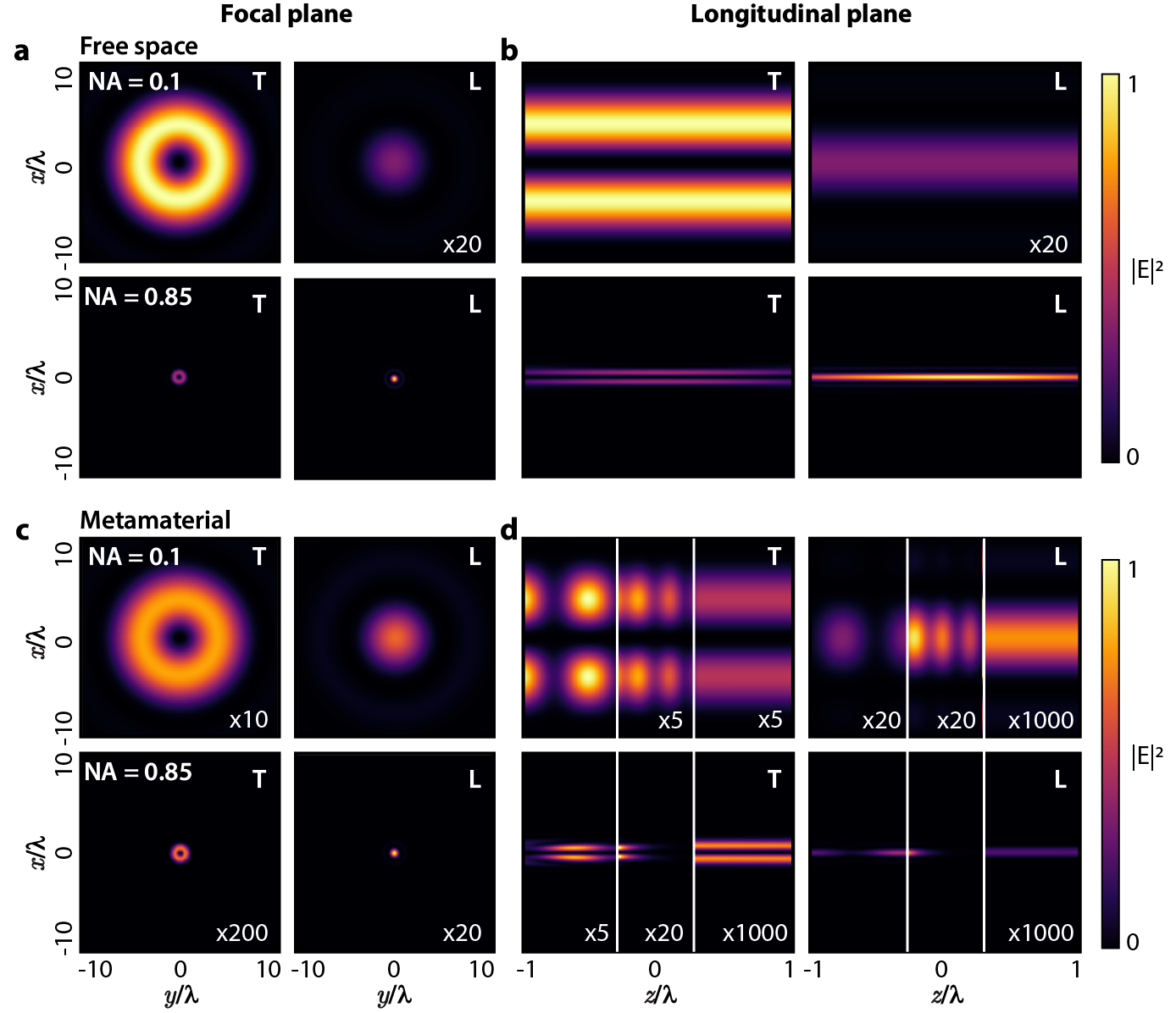}
    \caption{\textbf{Propagation and diffraction of a radial beam.} Intensity distributions of the transverse (T) and longitudinal (L) electric field components at a wavelength of $\lambda_{\rm ENZ}=660$~nm of a focused radial beam propagating through (a,b) free space and (c,d) the nanorod metamaterial. Cross sections are taken in (a,c) the focal plane and (b,d) across the 
    $xz$-plane in the case of (upper row of each panel) weak (NA=0.1) and~(lower row of each panel) strong (NA=0.85) focusing. The metamaterial parameters are as in Fig.~\ref{fig:Figure1}c).}
    \label{fig:FigS1}
\end{figure}

\begin{figure}
    \centering
    \includegraphics[width=\linewidth]{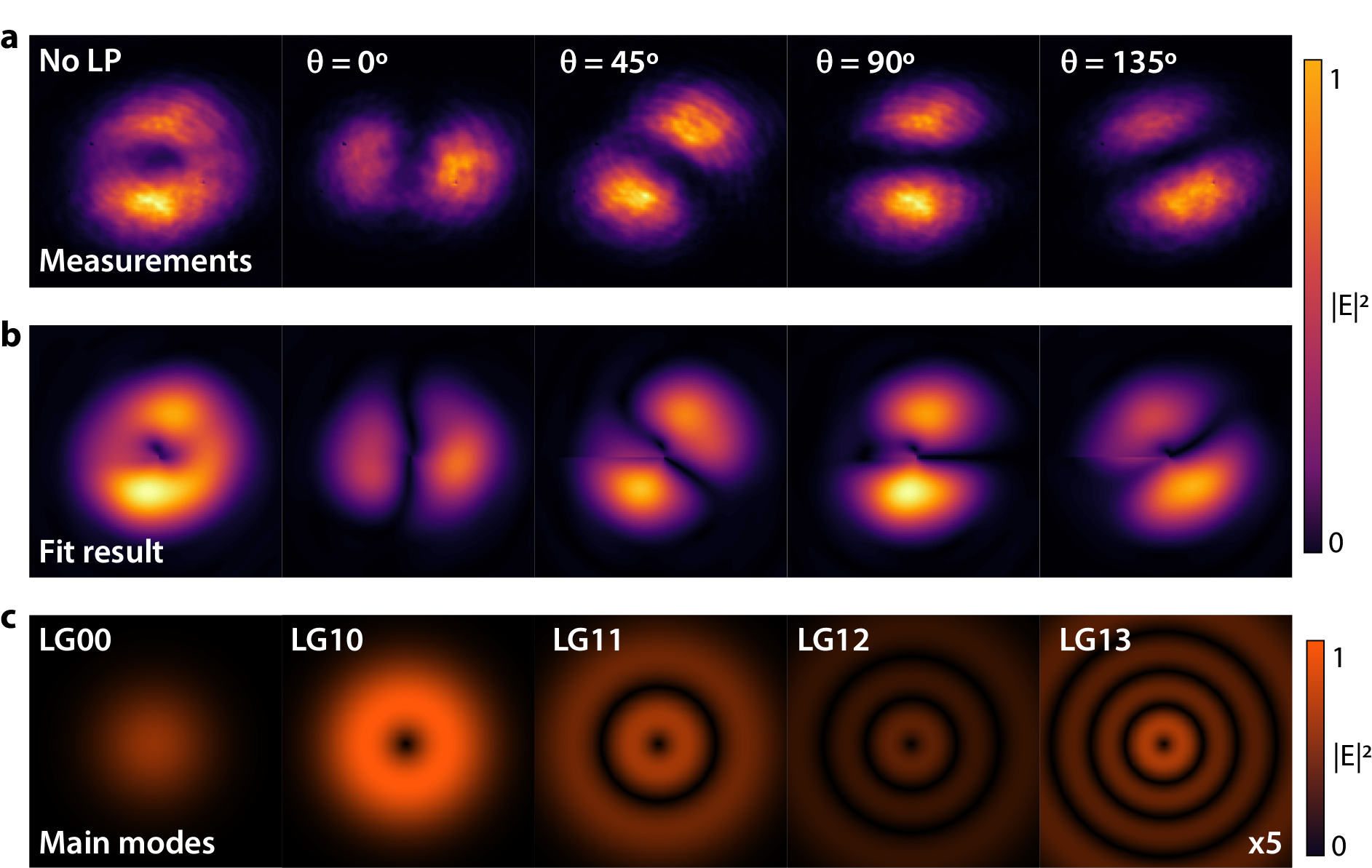}
    \caption{\textbf{Fitting of the modal content of a radial beam.} (a) Experimental intensity distributions of a radial beam propagating through the metamaterial and focused with an NA = 0.7 objective without and with a linear polariser for several axis orientations. (b) Intensity distributions obtained from the fitting procedure (Eq.~\ref{eq:e_fit}) corresponding to the conditions in (a). (c) Normalised intensities of the main LG modes contributing to the beam modal content, according to the fit results.}
    \label{fig:FigS2}
\end{figure}

\begin{figure}
    \centering
    \includegraphics[width=\linewidth]{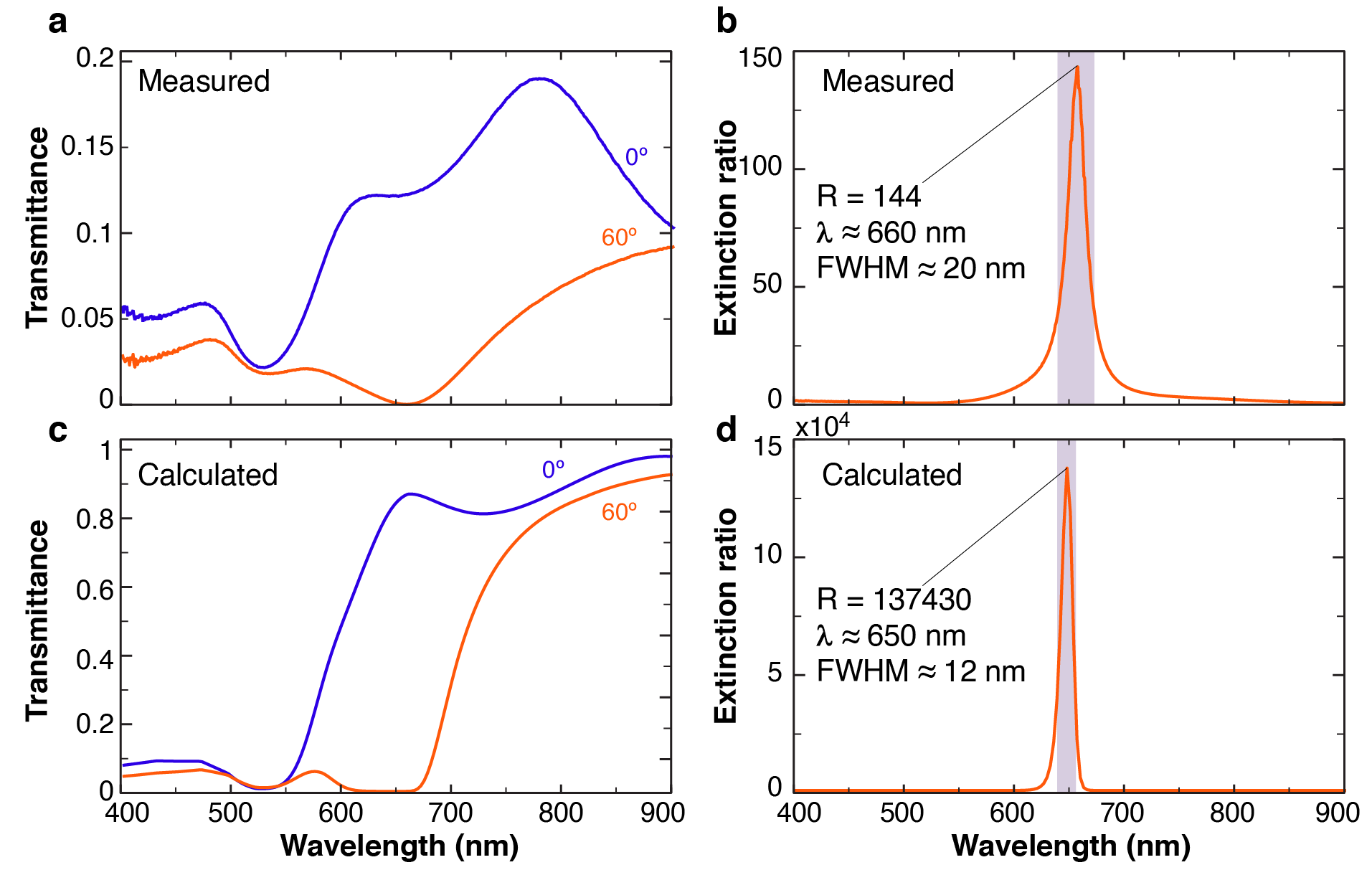}
    \caption{\textbf{Polarisation extinction spectroscopy.} (a,b) Measured and (c,d) calculated (a,b) transmission spectra and (c,d) polarisation extinction ratios (R) for $p$-polarised illumination at an angle of incidence of (blue) $0\degree$ and (orange) $60\degree$. The shaded areas in panels (b,d) show the bandwidth as the FWHM of the peak.}
    \label{fig:FigS3}
\end{figure}

\end{document}